\documentclass[aps,preprint]{revtex4-2}
\usepackage{graphicx}
\usepackage{lipsum}
\usepackage{float}
\usepackage{amsmath}
\usepackage{amssymb}
\usepackage{mathtools}
\include{mathmacros}

\begin{document}

\title{The structure of networks that evolve under a combination of \\
growth, via node addition and random attachment, \\
and contraction, via random node deletion  
}
\author{Barak Budnick, Ofer Biham and Eytan Katzav}
\affiliation{
Racah Institute of Physics, 
The Hebrew University, 
Jerusalem 9190401, Israel
}
\begin{abstract}

We present analytical results for the emerging structure
of networks that evolve via a combination of
growth (by node addition and random attachment) and 
contraction (by random node deletion).
To this end we consider a network model in which at each time step
a node addition and random attachment step takes place with
probability $P_{\rm add}$ and a random node deletion step 
takes place with probability $P_{\rm del}=1-P_{\rm add}$.
The balance between the growth and contraction processes is 
captured by the parameter $\eta = P_{\rm add}-P_{\rm del}$.
The case of pure network growth is described by $\eta=1$.
In case that $0 < \eta < 1$ the rate of node addition exceeds the rate of node deletion and
the overall process is of network growth.
In the opposite case, where $-1 < \eta < 0$,
the overall process is of network contraction,
while in the special case of $\eta=0$ the expected size
of the network remains fixed, apart from fluctuations.
Using the master equation and the generating function formalism   
we obtain a closed form expression for the 
time dependent degree distribution $P_t(k)$.
The degree distribution $P_t(k)$ includes a term that depends on the initial 
degree distribution $P_0(k)$, 
which decays as time evolves,
and an asymptotic distribution $P_{\rm st}(k)$ which 
is independent of the initial condition.
In the case of pure network growth ($\eta=1$)
the asymptotic distribution $P_{\rm st}(k)$ follows an exponential
distribution, while for $-1 < \eta < 1$ it consists of a sum of Poisson-like
terms and exhibits a Poisson-like tail.
In the case of overall network growth ($0 < \eta < 1$) the degree distribution $P_t(k)$ 
eventually converges to $P_{\rm st}(k)$. 
In the case of overall network contraction ($-1 < \eta < 0$)
we identify two different regimes. 
For $-1/3 < \eta < 0$ the degree distribution $P_t(k)$ quickly converges towards
$P_{\rm st}(k)$.
In contrast, for $-1 < \eta < -1/3$ 
the convergence of $P_t(k)$ is initially very slow and it gets
closer to $P_{\rm st}(k)$ only shortly before the network vanishes.
Thus, the model exhibits three phase transitions: a structural transition between two 
functional forms of $P_{\rm st}(k)$ at $\eta=1$, a transition between an
overall growth and overall contraction at $\eta=0$ and a dynamical transition
between fast and slow convergence towards $P_{\rm st}(k)$
at $\eta=-1/3$.
The analytical results are found to be in very good agreement
with the results obtained from computer simulations.

\end{abstract}

% 64.60.aq Networks
% 89.75.Da Systems obeying scaling laws

\maketitle

\newpage

\section{Introduction}

In the past 25 years or so, the field of network research has emerged as a major
field of study, which significantly contributed to the understanding of the structure and
dynamics of biological, social and technological networks
\cite{Dorogovtsev2003,Havlin2010,Newman2010,Estrada2011,Latora2017}.
It was found that empirical networks are typically 
small-world networks that exhibit
fat-tailed degree distributions with
scale free structures
\cite{Redner1998,Barabasi1999,Albert2002}. 
Much theoretical effort has focused on generic processes of network expansion or growth.
It was found that newly formed nodes tend to connect preferentially 
to nodes of high degree, and that this property leads to the
emergence of scale-free networks with power-law degree distributions
of the form
$P(k) \sim k^{- \gamma}$,
where $2 < \gamma \le 3$ 
and the second moment of the degree
distribution diverges 
\cite{Barabasi1999,Albert2002,Krapivsky2000,Dorogovtsev2000}.
In particular, the
Barab\'asi-Albert (BA) model 
exhibits a scale-free structure 
that emerges from the preferential-attachment process
\cite{Barabasi1999}.
In this model, at each 
time step a new node is added to the network and
forms links to $m$ of the existing nodes, such that the probability 
of an existing node
of degree $k$ to gain a link to the new node is proportional to $k$.
The degree distribution of the BA network exhibits a 
power-law tail
with $\gamma=3$.
Variants of the BA model were shown to yield power-law distributions with exponents in the range
$2 < \gamma \le 3$  
\cite{Krapivsky2000,Dorogovtsev2000,Bollobas2001}.
Another important class of network growth models is based on the duplication
of existing nodes,
where a new (daughter) node is connected to each neighbor of the
duplicated (mother) node with probability $p$, 
and in some cases it is also connected to the mother node itself 
\cite{Satorras2003,Chung2003,Krapivsky2005,Ispolatov2005,Ispolatov2005b,Bebek2006,Lambiotte2016,Bhat2016,Steinbock2017}.
The degree distributions of node duplication
networks follow a power-law distribution,
where $\gamma$ is a monotonically decreasing function of $p$ 
\cite{Chung2003,Ispolatov2005,Lambiotte2016,Bhat2016}.

The opposite scenario of network contraction has attracted increasing attention in recent years.
For example, the contraction processes of social networks was recently studied
\cite{Torok2017,Lorincz2019}.
Such networks may lose users due to loss of
interest, concerns about privacy or due to their migration to other social networks.
Another example is the evolution of gene networks, in which it was recently found that
the process of gene loss plays a significant role
\cite{Albalat2016}.
A different context of great practical importance is the cascading failure
of power-grids 
\cite{Daqing2014,Schafer2018},
in which the functional part of the network quickly contracts.
Infectious processes such as epidemics that spread in a network 
\cite{Satorras2001,Satorras2015}
lead to the contraction of the subnetwork of the susceptible (or uninfected) nodes,
and may thus be considered
as network contraction processes.
Similarly, network immunization schemes 
\cite{Satorras2002}
also belong to the class of network contraction
processes because they induce the contraction of the subnetwork of 
susceptible nodes.
The framework of network contraction is especially relevant in the
context of neurodegeneration, which is the progressive loss
of structure and function of neurons in the brain.
Such processes occur in 
normal aging 
\cite{Morrison1997}
as well as in
a large number of incurable neurodegenerative
diseases such as Alzheimer, Parkinson, Huntington and Amylotrophic
Lateral Sclerosis, which result in a gradual loss of cognitive and
motoric functions
\cite{Heemels2016}. 
These diseases differ in the specific brain regions or circuits
in which the degeneration occurs.
The analysis of the evolving structure 
may provide useful insight into 
the structural aspects of
the loss of neurons and synapses in
neurodegenerative processes
\cite{Arendt2015}.

Network contraction processes,  
which may result from inadvertent failures or from deliberate attacks,
were studied using the framework of percolation theory 
\cite{Albert2000,Cohen2000,Cohen2001,Gao2015,Yuan2015,Shao2015,Havlin2015,Shekhtman2015,Shekhtman2016,Yuan2016,Muro2016,Vaknin2017}.
It was shown that scale-free networks are resilient to attacks targeting
random nodes 
\cite{Cohen2000}, 
but are vulnerable to 
attacks that target high degree nodes or hubs 
\cite{Cohen2001}. 
In both cases, when
the number of deleted nodes 
exceeds some threshold, the network breaks down into disconnected components
\cite{Molloy1995,Molloy1998,Albert2000,Cohen2000,Cohen2001,Braunstein2016,Zdeborova2016}.
This analysis
provided important insights on the final stages of network collapse.
However, until recently the evolution of complex networks in the early and intermediate 
stages of their contraction process, before fragmentation, has not been 
studied in sufficient detail.
Understanding the patterns that emerge in the early and intermediate stages of network
failures or attacks is crucial for their detection and for devising ways to fix the
network and block such attacks. 

Recently we considered the evolution
of complex networks during generic contraction and collapse 
scenarios
\cite{Tishby2019,Tishby2020}.
These scenarios include 
random node deletion, preferential node deletion
and propagating node deletion.
The random node deletion process describes random failures or random 
attacks that do not target any specific type of nodes.
The process of preferential node deletion describes attacks that
preferentially target high degree nodes, while propagating node deletion
describes processes that propagate from an infected node
to its neighbors.
To analyze these processes we derived a master equation for
the time dependence of the degree distribution $P_t(k)$
in each one of the three network contraction scenarios.
In the scenario of random node deletion, the master equation is exact for
any ensemble of initial networks,
while in the scenarios of preferential and propagating node deletion it 
is exact for the case of configuration model networks, 
in which there are no degree-degree correlations
\cite{Newman2001,Catanzaro2005,Annibale2009,Roberts2011,Coolen2017}.
However, it was shown to
provide reasonably accurate results for the time-dependent degree
distributions even in networks that exhibit degree-degree correlations.
Using the master equation we established that when
networks contract via any of the node deletion scenarios described above,
their degree distributions evolve towards a Poisson distribution,
namely they become Erd{\H o}s-R\'enyi (ER) networks
\cite{Erdos1959,Erdos1960,Erdos1961}.
These networks belong to an ensemble of maximum entropy 
random graphs
\cite{Coolen2017}.

The emerging structure of networks that evolve under a combination 
of growth and contraction processes
was studied in Refs.
\cite{Moore2006,Bauke2011,Ghoshal2013}.
These papers focus on 
the regime in which the 
overall process is of network growth.
A particularly interesting case is of networks that grow via a 
combination of preferential attachment and random attachment,
which exhibit a degree distribution with a power-law tail.
It was found that under low rate of random node deletion
the degree distribution maintains its power-law tail.
However, above some threshold (that depends on the mixture of random
attachment and preferential attachment) the power-law tail is lost and 
is replaced by a discrete exponential degree distribution 
(which is also known as a geometric distribution).
The phase boundary between the two phases was calculated
(using different parameterizations), giving rise to highly insightful
phase diagrams 
\cite{Bauke2011,Ghoshal2013}.
The combination of growth via node addition and random attachment and
contraction via random node deletion was also studied
\cite{Moore2006}.
In the limit of pure growth this
model gives rise to networks that exhibit an exponential (geometric) degree distribution
\cite{Moore2006,Steinbock2017}.
As mentioned above, 
Refs. \cite{Moore2006,Bauke2011,Ghoshal2013} 
focus on the steady state solution of the degree distribution in
case that the overall process is of network growth.
The complementary regime in which the rate of node deletion exceeds
the rate of node addition has not been studied. 

In this paper we analyze the emerging structure of networks that evolve
under a combination of growth (via node addition and random attachment)
and contraction (via random node deletion).
We derive a master equation for
the time dependence of the degree distribution
under this combination of growth and contraction processes.
Using the generating function formalism  
we obtain a closed form expression for the degree distribution $P_t(k)$.
It includes a term that depends on the initial condition, 
which decays as time evolves,
and an asymptotic term which is an attractive fixed point.
We identify a phase transition 
between the phase of pure network growth and the phase that combines
growth and contraction.
This transition implies that even the slightest rate of node deletion leads
to a qualitative change in the nature of the degree distribution.
In the regime of overall network growth, $P_t(k)$ eventually converges towards
the asymptotic steady state form $P_{\rm st}(k)$.
In contrast, in the regime of overall network contraction  
the asymptotic degree distribution is not
always reached due to the finite life-time of the network.
This gives rise to a second phase transition, between the phase of 
overall network growth and the phase of overall network contraction.
In the phase of overall network contraction we identify a third transition,
between the case of low deletion rate, 
in which the degree distribution $P_t(k)$ quickly approaches
$P_{\rm st}(k)$,
and the case of high deletion rate, in which
the convergence of $P_t(k)$ is initially very slow and it gets
closer to $P_{\rm st}(k)$ only shortly before the network vanishes.
The analytical results are found to be in very good agreement
with the results obtained from computer simulations.

The paper is organized as follows. 
In Sec. II we describe the dynamical model that
combines growth (via node addition and random attachment) and
contraction (via random node deletion).
In Sec. III we derive a master equation for the time dependent
degree distribution $P_t(k)$.
In Sec. IV we use the master equation to derive a differential equation 
for the generating function $G_t(u)$
of the degree distribution and present its time-dependent solution.
In Sec. V we present a closed-form expression for the degree distribution $P_t(k)$,
obtained from $G_t(u)$.
In Sec. VI we calculate the mean and variance of the degree distribution.
%present the solution of the master equation and
%analyze its properties.
The results are summarized and discussed in Sec. VII.
In Appendix A we solve the differential equation for $G_t(u)$ and
%we calculate the time dependent generating function $G_t(u)$
%associated with the master equation and
extract the degree distribution $P_t(k)$.
In Appendix B we calculate the degree distribution $P_t(k)$
in the special case of pure network growth.

\section{The model}

Consider a network that evolves as follows.
At each time step, one of two possible processes takes place:
(a) growth step: with probability $P_{\rm add}$ an isolated node (of degree $k=0$)
is added to the network. The node addition is followed by the addition of $m$ edges between pairs 
of random nodes (which have not been connected before).
This is done by repeating the following step $m$ times:
each time two random nodes (which have not been connected before)
are selected and are connected to each other by an edge;
(b) contraction step: with probability
$P_{\rm del}=1-P_{\rm add}$ a random node is deleted, together with its edges.

When a growth step is selected at time $t$, the network size increases
according to $N_{t+1}=N_t + 1$, while the degrees of the $m$ pairs
of newly connected nodes increase from $k_i$ to $k_i+1$.
When a contraction step is selected at time $t$, the network size
decreases according to $N_{t+1}=N_t - 1$.
Consider a node of degree $k$, whose neighbors are of
degrees $k'_r$, $r=1,2,\dots,k$.
Upon deletion of such node 
the degrees of its neighbors are reduced to
$k'_r-1$, $r=1,2,\dots,k$. 

We denote the initial number of nodes in the network at time $t=0$ by $N_0$.
The expectation value of the number of nodes in the network at time $t$ is
\begin{equation}
N_t = N_0 + \eta t,
\label{eq:Nt}
\end{equation}

\noindent
where
\begin{equation}
\eta = P_{\rm add} - P_{\rm del}.
\end{equation}

\noindent
The parameter $\eta$ provides a convenient classification of the possible scenarios.
The case of pure growth is described by $\eta=1$.
For $0 <\eta < 1$ the overall process is of network growth, 
while for $-1 \le \eta <0$ the overall
process is of network contraction.
In the special case of $\eta=0$ the network size remains the same, apart from 
possible fluctuations.
It is convenient to express the probabilities $P_{\rm add}$ and $P_{\rm del}$ 
in terms of the parameter $\eta$, namely

\begin{equation}
P_{\rm add} =  \frac{1+\eta}{2} 
\label{eq:Padd}
\end{equation}

\noindent
and

\begin{equation}
P_{\rm del} = \frac{1-\eta}{2}.
\label{eq:Pdel}
\end{equation}

\noindent
In the case of $-1 < \eta < 0$ it is convenient to define the normalized time variable

\begin{equation}
\tau = \frac{ | \eta | t }{N_0},
\label{eq:tau}
\end{equation}

\noindent
that measures the fraction of nodes that are deleted from the network
up to time $t$. The expected size of the contracting network at time $t$ 
can be expressed by $N_t = N_0(1-\tau)$.
Note that the network vanishes at $\tau=1$.

In the model considered here the $m$ edges added at time $t$ 
connect pairs of existing random nodes.
This model is different from the random attachment model studied in Ref.
\cite{Moore2006}, in which the new edges connect the new node
to $m$ random nodes in the network.
Thus, in the model of Ref. \cite{Moore2006}
the degree of the new node upon its addition to the
network is $k=m$. 
As a result, the degree distribution exhibits a cusp at $k=m$,
separating between the regime of low degrees, $k<m$, and
the regime of high degrees, $k>m$.
In the model studied here the new node is added with degree $k=0$
and gains links one at a time in subsequent time steps.
As a result, the degree distribution 
exhibits the same functional form over the whole range of
possible values of $k$.
In that sense, the model studied here is somewhat simpler, while fundamentally belonging
to the same class of random attachment models.

\section{The master equation}

Consider an ensemble of 
networks of size $N_0$ at time $t=0$,
whose initial degree distribution is given by $P_0(k)$.
The networks evolve under a combination of growth 
(via node addition and random attachment)
and contraction (via random node deletion).
Below we derive a master equation 
\cite{Vankampen2007,Gardiner2004}
that describes the time
evolution of the degree distribution 

\begin{equation}
P_t(k) = \frac{N_t(k)}{N_t},
\label{eq:P_t(k)}
\end{equation}

\noindent
where $N_t(k)$, $k=0,1,\dots$, is the number of nodes of degree $k$ at time $t$
and $N_t = \sum_k N_t(k)$ is the network size at time $t$.
The master equation formulation was used before in network growth processes
\cite{Krapivsky2000,Dorogovtsev2000}
and in processes that combine growth and contraction
\cite{Moore2006,Bauke2011,Ghoshal2013}.

In general, the master equation accounts for the time 
evolution of the degree distribution $P_t(k)$ over an ensemble of networks
of the same initial size $N_0$ and initial degree distribution $P_0(k)$, which
are exposed to the same dynamical processes.
In order to derive the master equation, we first consider the
time evolution of $N_t(k)$, which can be expressed in terms
of the forward difference

\begin{equation}
\Delta_t N_t(k) = N_{t+1}(k) - N_t(k).
\end{equation}

\noindent
In the case of a growth step, the
addition of an isolated node increases by $1$ the number of nodes of degree $k=0$,
namely $N_t(0) \rightarrow N_t(0)+1$.
The contribution of this process to the evolution of $N_t(k)$ is given by

\begin{equation}
A_t(k) = 
P_{\rm add} \ \delta_{k,0},
\label{eq:Qk}
\end{equation}

\noindent
where $\delta_{i,j}$ is the Kronecker delta symbol.
The probability that a random node of degree $k$ will gain an additional edge at
time $t$ is given by

\begin{equation}
U_t(k \rightarrow k+1) = 2m P_{\rm add}  \frac{N_t(k)}{N_t}.
\label{eq:Uk}
\end{equation}

\noindent
Similarly, the probability that a random node of degree $k-1$ will gain an 
additional edge is

\begin{equation}
U_t(k-1 \rightarrow k) = 2m P_{\rm add} \frac{N_t(k-1)}{N_t}.
\label{eq:Uk+1}
\end{equation}

\noindent
Here we use the convention that
$N_t(-1)=0$.

In the case of a contraction step,
the probability that the node 
selected for deletion at time $t$ is of degree $k$ is
given by $N_t(k)/N_t$.
Thus, the rate of change of $N_t(k)$  due to a deletion of a node of degree $k$
is given by

\begin{equation}
D_t(k) = 
- P_{\rm del} \frac{N_t(k)}{N_t}.
\label{eq:Rk}
\end{equation}

\noindent
Consider the case in which the process that takes place at time $t$ 
is the deletion of a random node.
In case that the deleted node is of degree $k'$,
it affects $k'$ adjacent nodes, which lose one link each. 
The probability of each one of these $k'$ nodes
to be of degree $k$ is given by
$k N_t(k)/[ N_t \langle K\rangle_t ]$, where
$\langle K \rangle_t$ is the mean degree.
We denote by $W_t(k \rightarrow k-1)$ the expectation value of
the number of nodes of degree $k$ that lose a link at time $t$ and
are reduced to degree $k-1$.
Summing up over all possible values of $k'$,  
we find that the effect of node deletion on neighboring nodes
of degree $k$ is given by

\begin{equation}
W_t(k \rightarrow k-1) = P_{\rm del} \frac{kN_t(k)}{N_t}.
\label{eq:Wk}
\end{equation}

\noindent
Similarly, the effect on neighboring nodes of degree $k+1$ 
accounts to

\begin{equation}
W_t(k+1 \rightarrow k) = P_{\rm del} \frac{ (k+1)N_t(k+1)}{N_t}.
\label{eq:Wk+1}
\end{equation}

\noindent
Combining the  
effects on the time dependence of $N_t(k)$ 
we obtain

\begin{eqnarray}
\Delta_t N_t(k) &=&
A_t(k ) 
+ \left[ U_t(k-1 \rightarrow k) - U_t(k \rightarrow k+1) \right]
\nonumber \\
&+& D_t(k) 
+  \left[ W_t(k+1 \rightarrow k) - W_t(k \rightarrow k-1) \right].
\label{eq:RWW}
\end{eqnarray}

\noindent
Inserting the expressions for 
$A_t(k)$, $D_t(k)$, 
$U_t(k - 1 \rightarrow k)$,
$U_t(k \rightarrow k+1)$,
$W_t(k \rightarrow k-1)$ and
$W_t(k+1 \rightarrow k)$,
from Eqs.
(\ref{eq:Qk}),
(\ref{eq:Rk}),
(\ref{eq:Uk}), (\ref{eq:Uk+1}), (\ref{eq:Wk}) and (\ref{eq:Wk+1}), respectively,
we obtain

\begin{eqnarray}
\Delta_t N_t(k) &=&
P_{\rm add} \left[ \delta_{k,0} + 2m \frac{  N_t(k-1) - N_t(k)  }{N_t} \right]
\nonumber \\
&+&
P_{\rm del} \frac{(k+1)[ N_t(k+1) - N_t(k) ]}{N_t}.
\label{eq:DeltaNtk2}
\end{eqnarray}

Since nodes are discrete entities the processes of node addition and 
deletion are intrinsically discrete. Therefore, the replacement of the
forward difference $\Delta_t N_t(k)$ by a time derivative of the form
$dN_t(k)/dt$ involves an approximation.
The error associated with this approximation was shown to be
of order $1/N_t^2$, which quickly vanishes for sufficiently large networks
\cite{Tishby2019}.
Therefore, the difference equation (\ref{eq:DeltaNtk2}) can be replaced by the
differential equation

\begin{eqnarray}
\frac{d}{dt} N_t(k) &=&
P_{\rm add} \left[ \delta_{k,0} + 2m \frac{  N_t(k-1) - N_t(k)  }{N_t} \right]
\nonumber \\
&+&
P_{\rm del}
\frac{(k+1)[ N_t(k+1) -  N_t(k) ]}{N_t}.
\label{eq:DeltaNtkde0}
\end{eqnarray}

\noindent
The derivation of the master equation is completed by taking the
time derivative of Eq. (\ref{eq:P_t(k)}), which is given by

\begin{equation}
\frac{d}{dt} P_t(k) = 
\frac{1}{N_t} \frac{d}{dt} N_t(k) - \frac{N_t(k)}{N_t^2} \frac{d}{dt} N_t.
\label{eq:dPt_Nt}
\end{equation}

\noindent
Inserting the time derivative of $N_t(k)$ from Eq. (\ref{eq:DeltaNtkde0}) 
and using the fact that
$d N_t/dt=\eta$ 
[from Eq. (\ref{eq:Nt})],
we obtain the following master equation 

\begin{eqnarray}
\frac{d}{dt} P_t(k) &=&
\frac{1+\eta}{2 N_t} [ \delta_{k,0} - P_t(k) ]
+ \frac{ m (1+\eta) }{N_t} [P_t(k-1)-P_t(k)]
\nonumber \\
&+&
\frac{1-\eta}{2 N_t}
\left[ (k+1)P_t(k+1) - k P_t(k) \right],
\label{eq:dP(t)/dtRC0}
\end{eqnarray}

\noindent
where we have also expressed $P_{\rm add}$ and $P_{\rm del}$
in terms of $\eta$, using Eqs. (\ref{eq:Padd}) and (\ref{eq:Pdel}).
In essence, the master equation consists of a set of coupled ordinary differential equations
for $P_t(k)$, $k=0,1,2,\dots$.
In Eq. (\ref{eq:dP(t)/dtRC0}) we use the convention that
$P_t(-1)=0$.
For a given initial size $N_0$ and initial degree distribution $P_0(k)$,
the master equation can be solved by direct numerical integration.

In the case of pure growth ($\eta=1$) the master equation is reduced to
the form

\begin{equation}
\frac{d}{dt} P_t(k) =
\frac{1}{ N_t} [ \delta_{k,0} - P_t(k) ]
+ \frac{ 2 m }{N_t} [P_t(k-1)-P_t(k)].
\label{eq:dP(t)/dtRC01}
\end{equation}

\section{The generating function}

Below we solve the master equation using the generating function formalism.
We denote the generating function by

\begin{equation}
G_t(u) = \sum_{k=0}^{\infty} u^k P_t(k),
\end{equation}

\noindent
which is the Z-transform of the degree distribution $P_t(k)$
\cite{Phillips2015}.
Multiplying Eq. (\ref{eq:dP(t)/dtRC0}) by $u^k$ and summing up over $k$,
we obtain a partial differential equation for the generating function,
which is given by

\begin{equation}
N_0 \left(1 + \frac{\eta t}{N_0} \right) \frac{ \partial G_t(u) }{\partial t} 
- \frac{ 1 - \eta }{2} (1-u) \frac{ \partial G_t(u) }{\partial u}
+ \frac{1+\eta}{2} \left[ 2m(1-u) + 1 \right] G_t(u) 
= \frac{1+\eta}{2}.
\label{eq:diffeq0a}
\end{equation}

\noindent
This is a first order inhomogeneous linear partial differential equation of two variables.
Note that $\eta=1$ is a singular point of this differential equation.
At $\eta=1$ the coefficient of the term that includes the 
derivative of $G_t(u)$ with respect to $u$ vanishes,
thus reducing the order of the equation.
This is reflected in the fact that for $\eta=1$ the steady-state solution of Eq. (\ref{eq:diffeq0a})
is of a different nature than the solution for $-1 < \eta <1$,
implying a structural phase transition at $\eta=1$.

For the analysis of Eq. (\ref{eq:diffeq0a}) it is useful to
define the parameter

\begin{equation}
r = \frac{1+\eta}{1-\eta}.
\label{eq:r}
\end{equation}

\noindent
In the regime of overall network growth, in which $0 < \eta < 1$,
the parameter $r$ is a monotonically increasing function of $\eta$,
which rises from $r=1$ for $\eta=0$ to $r \rightarrow \infty$ at
$\eta \rightarrow 1$.
In the regime of overall network contraction, where $-1 < \eta < 0$,
$r$ is also a monotonically increasing function of $\eta$, which rises
from $r=0$ at $\eta=-1$ to $r=1$ at $\eta=0$.

In Appendix A
we use the method of characteristics to solve Eq. (\ref{eq:diffeq0a})
and obtain the generating function $G_t(u)$ for
$-1 \le \eta < 1$.
It is given by

\begin{eqnarray}
G_t(u) &=& \alpha_t^r e^{-2rm(1-\alpha_t)(1-u)} G_0[1-\alpha_t(1-u)]
\nonumber \\
&+& r \int_{\alpha_t}^{1} y^{r-1} e^{-2rm(1-u)(1-y)} dy,
\label{eq:Gtu20}
\end{eqnarray}

\noindent
where $G_0(x)$ is the generating function of the initial degree distribution $P_0(k)$ 
and

\begin{equation}
\alpha_t = 
\left\{
\begin{array}{ll}
  \left( 1 +  \frac{\eta t}{N_0} \right)^{ - \frac{ 1-\eta }{2 \eta} }     & 0 < \eta < 1 \\
\exp \left( - \frac{ t }{2N_0} \right) &  \eta = 0 \\ 
\left( 1 -   \frac{| \eta | t}{N_0} \right)^{ \frac{ 1 + |\eta| }{2 |\eta|} }    & -1 \le \eta < 0.
\end{array}
\right.  
\label{eq:alpha_t}
\end{equation}

The generating function $G_t(u)$, given by Eq. (\ref{eq:Gtu20}),
consists of two terms.
The first term depends on the degree distribution of the initial network
while the second term does not depend on the properties of the initial network.
Note that $G_t(1)=1$, reflecting the normalization of the distribution $P_t(k)$.
Plugging $u=1$ in the first term of Eq. (\ref{eq:Gtu20}) shows that
the weight of the first term is equal to 

\begin{equation}
w_t = \alpha_t^r,
\label{eq:w_t}
\end{equation}

\noindent
where $\alpha_t$ decreases monotonically as time evolves
(from its initial value of $\alpha_0=1$).
Therefore, the decay of $w_t$ as time evolves controls the 
rate at which the information about the initial network structure is lost.

Note that in Eq. (\ref{eq:alpha_t}) 
the expression $\alpha_t = (1+\eta t/N_0)^{ - \frac{1-\eta}{2\eta} }$
is valid for any $\eta \ne 0$. 
However, there is a qualitative difference in the behavior of $\alpha_t$
between the regime of overall network growth ($\eta>0$) and the regime 
of overall network contraction ($\eta<0$).
This difference is emphasized by the presentation of Eq. (\ref{eq:alpha_t}),
where we express it somewhat differently in the two regimes.
More specifically, in the regime of overall network growth 
the parameter $\alpha_t$ gradually decreases
towards zero as time evolves and the network
continues to grow for an unlimited period of time.
In contrast, in the regime of overall network contraction, 
$\alpha_t$ reaches zero after a finite time, namely at

\begin{equation}
t_{\rm vanish}= \frac{ N_0 }{ | \eta |},
\label{eq:t_vanish}
\end{equation}

\noindent
which is the time it takes for the
network to vanish completely.

\begin{figure}
\begin{center}
\includegraphics[width=6cm]{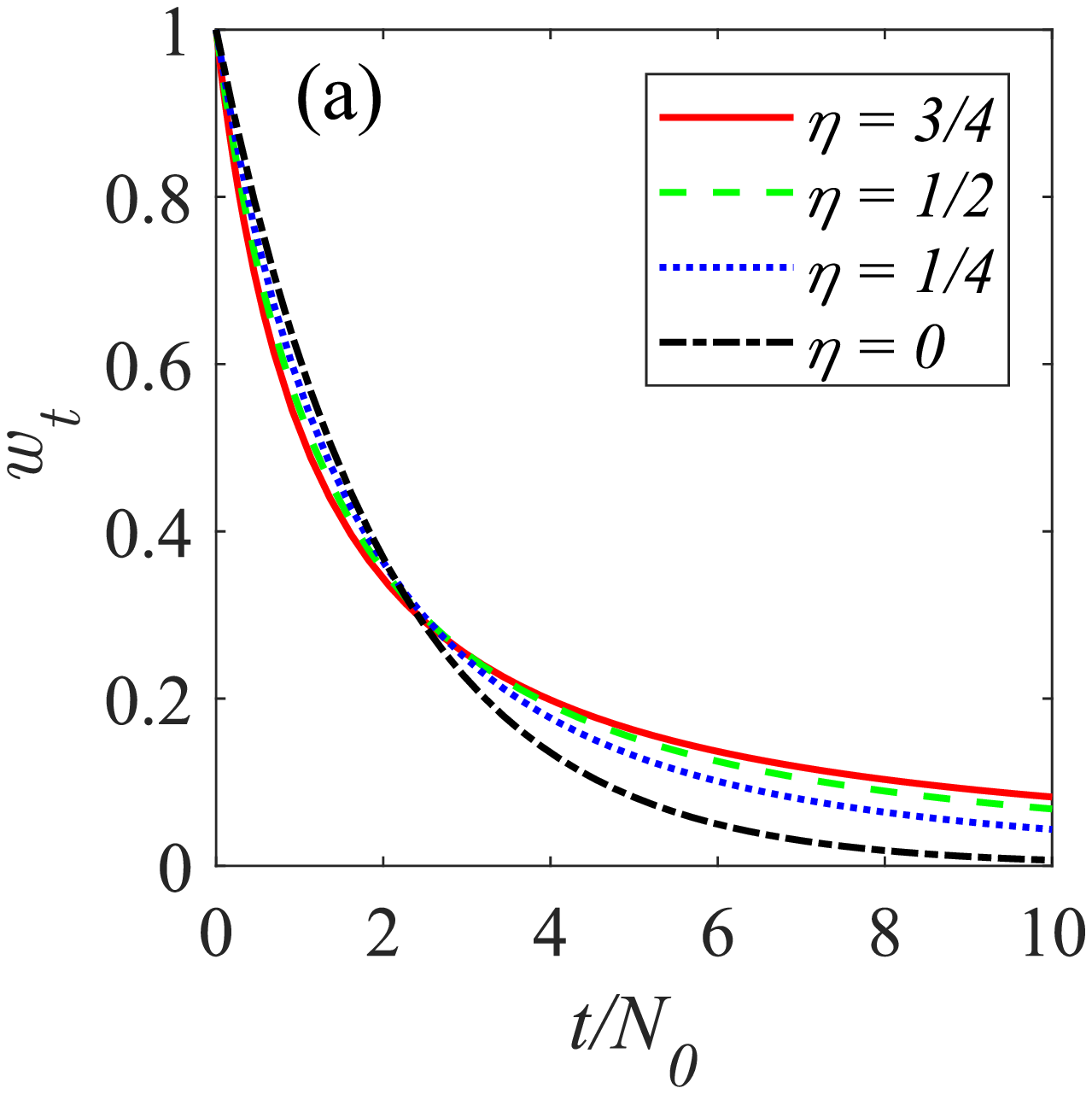} 
\\
\includegraphics[width=6cm]{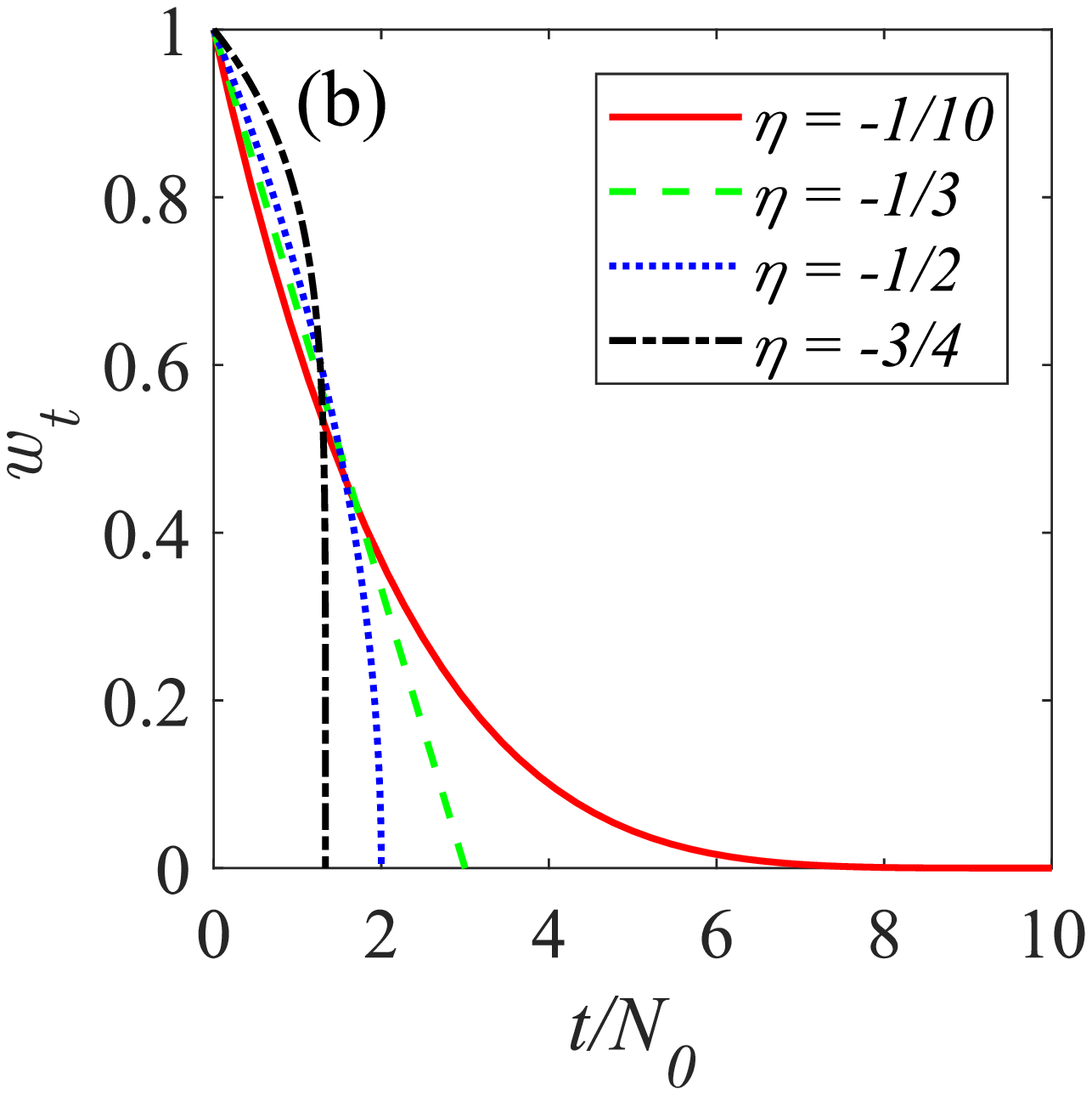}
\end{center}
\caption{
(Color online)
The coefficient $w_t$ as a function of $t/N_0$ for
networks that evolve under a combination
of growth via random node addition and random attachment and
contraction via random node deletion 
for (a) $0 \le \eta < 1$; and (b) $-1 < \eta < 0$,
obtained from Eqs. (\ref{eq:alpha_t})-(\ref{eq:w_t}),
where $r$ is given by Eq. (\ref{eq:r}).
In case that $\eta \ge 0$ the coefficient $w_t$
decreases monotonically as a function of $t$
but converges towards $0$ only asymptotically.
In case that $\eta<0$ the coefficient $w_t$ vanishes at 
a finite time $t_{\rm vanish}=N_0/| \eta |$.
The curve of $w_t$ vs. $t/N_0$ is
convex for $-1/3 < \eta < 0$ and concave
for $-1 < \eta < -1/3$.
}
\label{fig:1}
\end{figure}

In Fig. \ref{fig:1} we present the
coefficient $w_t$ as a function of $t/N_0$ for
networks that evolve under a combination
of growth (via random node addition and random attachment) and
contraction (via random node deletion)
for (a) $0 \le \eta < 1$; and (b) $-1 < \eta < 0$,
obtained from Eq. (\ref{eq:alpha_t}),
where $r$ is given by Eq. (\ref{eq:r}).
In case that $\eta \ge 0$ the coefficient $w_t$
decreases monotonically as a function of $t$
but converges towards $0$ only asymptotically.
In case that $\eta<0$, the coefficient $w_t$ vanishes after
a finite time $t_{\rm vanish}$, given by Eq. (\ref{eq:t_vanish}).

For $-1 < \eta < 0$ the weight $w_t$ can be expressed in the form

\begin{equation}
w_t = \left( 1 - \frac{t}{t_{\rm vanish}} \right)^{ \frac{1-|\eta|}{2 |\eta|} }.
\label{eq:wt}
\end{equation}

\noindent
In this range the time derivative of $w_t$ 
is given by

\begin{equation}
\frac{d w_t}{dt} = 
- \frac{1-|\eta|}{ 2 |\eta| t_{\rm vanish} }
\left( 1 - \frac{t}{t_{\rm vanish}} \right)^{ \frac{1-3|\eta|}{2 |\eta|} }.
\label{eq:dwtdt}
\end{equation}

\noindent
This derivative represents the rate at which the memory of the
initial network is lost.
For $-1/3 < \eta < 0$ the exponent in Eq. (\ref{eq:dwtdt}) is positive, while
for $-1 < \eta < -1/3$ it is negative.
Therefore, as $\eta$ crosses $-1/3$ the derivative
$d w_t/dt |_{t=t_{\rm vanish}}$ changes discontinuously
from $0$ to $-\infty$.
Such discontinuous changes represent a typical behavior at a phase transition.

\begin{figure}
\begin{center}
\includegraphics[width=6cm]{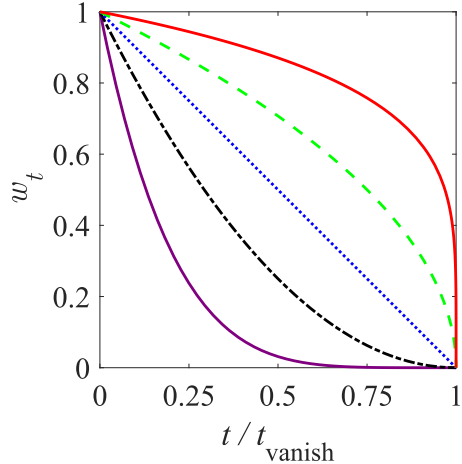} 
\end{center}
\caption{
(Color online)
The coefficient $w_t$ as a function of $t/t_{\rm vanish}$ for
networks that evolve under a combination
of growth via random node addition and random attachment and
contraction via random node deletion 
for $\eta = -1/11$, $-1/5$, $-1/3$, $-1/2$ and $-5/7$
(from left to right),
obtained from Eq. (\ref{eq:wt}),
which is valid for $\eta<0$.
The curve of $w_t$ vs. $t/N_0$ is
convex for $-1/3 < \eta < 0$ and concave
for $-1 < \eta < -1/3$,
while for $\eta=-1/3$ it follows a straight line.
}
\label{fig:2}
\end{figure}

In Fig. \ref{fig:2} we present the
coefficient $w_t$ as a function of $t/t_{\rm vanish}$ for
networks that evolve under a combination
of growth (via random node addition and random attachment) and
contraction (via random node deletion)
for $-1 < \eta < 0$.
As $t \rightarrow t_{\rm vanish}$ the slope
$d w_t/dt$ vanishes for $-1/3 < \eta < 0$ and
diverges for $-1 < \eta < -1/3$.

As time evolves, the first term in 
Eq. (\ref{eq:Gtu20}) decreases while the second term
increases and flows towards an asymptotic state,
given by

\begin{equation}
G_{\rm st}(u)  =  
r \int_{0}^{1} y^{r-1} e^{-2rm(1-u)(1-y)} dy.
\label{eq:Gtu20st0}
\end{equation}

\noindent
Expressing the integral in terms of the 
lower incomplete gamma function
$\gamma(s,x)$, given by Eq. (\ref{eq:lowergamma}) in Appendix A,
we obtain 

\begin{equation}
G_{\rm st}(u)  =   
r e^{-2 r m(1-u) } [-2 r m (1-u) ]^{-r} 
\gamma[r,-2rm(1-u)].  
\label{eq:Gtu20st}
\end{equation}

\noindent
Using this notation, one can express Eq. (\ref{eq:Gtu20}) in the
form

\begin{eqnarray}
G_t(u) &=& \alpha_t^r e^{-2rm(1-\alpha_t)(1-u)} G_0[1-\alpha_t(1-u)]
\nonumber \\
&+&  
\left\{ 1 - \frac{ \gamma[r,-2rm \alpha_t (1-u) ] }{ \gamma[r,-2rm(1-u)]  }   \right\}
G_{\rm st}(u),
\label{eq:Gtu20b}
\end{eqnarray}

\noindent
where the first term captures the memory of the degree distribution 
of the initial network while the second term includes the components
that do not depend on the initial degree distribution.
As time evolves, the first term decays while the second term 
converges towards the asymptotic form, given by Eq. (\ref{eq:Gtu20st}).

\section{The degree distribution}

In Appendix A we extract the time dependent degree distribution $P_t(k)$
from the generating function $G_t(u)$. 
It is given by

\begin{eqnarray}
P_t(k) &=& \alpha_t^r \frac{ e^{-2 r m(1-\alpha_t) }}{k!}
\sum_{i=0}^k \binom{k}{i} \alpha_t^i 
\frac{d^i G_0(u)}{du^i} \bigg\vert_{u=1-\alpha_t}
\left[ 2 r m(1-\alpha_t) \right]^{k-i}  
\nonumber \\
&+& r e^{-2rm} \frac{(2rm)^k}{k!} \int_{\alpha_t}^{1}
y^{r-1} e^{2rmy} (1-y)^k dy.
\label{eq:Ptk0}
\end{eqnarray}

The dependence of $P_t(k)$ on the initial degree distribution $P_0(k)$ 
is captured by first term of Eq. (\ref{eq:Ptk0}), while the second term 
is an asymptotic solution that does not depend on the initial condition.
This asymptotic solution is essentially an attractive fixed point.
The rate of convergence depends on the parameter $\eta$.
More precisely, it is regulated by the coefficient
$w_t = \alpha_t^r$ which appears in front of the term that 
captures the initial condition. 
As mentioned in the previous section, the dependence of $w_t$ 
on time is different in the regime
of overall network growth ($\eta>0$) 
and the regime of overall network contraction ($\eta<0$).
For $\eta>0$ the coefficient $w_t$ decays asymptotically like

\begin{equation}
w_t \sim t^{- \frac{r}{r-1} }.
\end{equation}

\noindent
Thus, for sufficiently long times the memory of the initial degree
distribution is completely lost and $P_t(k)$ approaches its asymptotic form.

In the case of $\eta<0$ the coefficient $w_t$ decays as time
evolves until it vanishes at a finite time $t_{\rm vanish}$.
At the point $\eta=-1/3$ there is transition from a convex shape of
$w_t$ as a function of the time $t$
(for $-1/3 < \eta < 0$) to a concave shape (for $-1 \le \eta < -1/3$),
as can be seen in Fig. \ref{fig:2}.
For $\eta > -1/3$, as $t \rightarrow t_{\rm vanish}$ the derivative
$d w_t/dt \rightarrow 0$.
In contrast, for $\eta < - 1/3$,
as $t \rightarrow t_{\rm vanish}$ the derivative
$d w_t/dt \rightarrow - \infty$.
This sharp discontinuity 
in $d w_t/dt \vert_{t_{\rm vanish}}$ at $\eta=-1/3$
pinpoints the location of the dynamical transition.
Note that the value of $\eta=-1/3$ corresponds to the situation in which
$P_{\rm add}=1/3$ and $P_{\rm del}=2/3$,
namely on average there are two node deletion steps for each 
node addition step.

From Eq. (\ref{eq:Ptk0}) one observes that 
on top of the overall dependence on $w_t$,
the rate of convergence 
of $P_t(k)$ towards its asymptotic value depends on the degree $k$.
The asymptotic form of $P_t(k)$ in the long time limit can be obtained
by inserting $\alpha_t=0$ in Eq. (\ref{eq:Ptk0}).
It yields

\begin{equation}
P_{\rm st}(k) =
r e^{-2rm} \frac{(2rm)^k}{k!} \int_{0}^{1}
y^{r-1} e^{2rmy} (1-y)^k dy.
\label{eq:Pstk}
\end{equation}

\noindent
The right hand side of Eq. (\ref{eq:Pstk}) can be expressed 
in the form

\begin{equation}
P_{\rm st}(k) = 
e^{-2rm} \frac{(2rm)^k}{k!}
r B(k+1,r)
\  _1 F_1 \left( 
\begin{array}{c}
r \\
k+r+1
\end{array}
\bigg\vert 2rm \right),
\label{eq:Pstkg}
\end{equation}

\noindent
where $B(m,n)$ is the beta function
and $_1 F_1(\cdot)$ is the confluent hypergeometric function
\cite{Olver2010}.

The tail of the steady state degree distribution $P_{\rm st}(k)$, 
where $k \gg r$ can be reduced to

\begin{equation}
P_{\rm st}(k) \simeq \Gamma(r+1) k^{-r} e^{-2rm} \frac{(2rm)^k}{k!}.
\label{eq:Pktail}
\end{equation}

\noindent
This tail resembles the Poisson distribution in the sense that it
satisfies the condition that $P_{\rm st}(k)/P_{\rm st}(k-1) \propto 1/k$.

In the special case of $\eta=0$ (where $r=1$),
which represents a perfect balance between the growth and
contraction processes, the distribution
$P_{\rm st}(k)$
takes a particularly simple form

\begin{equation}
P_{\rm st}(k; \eta=0) = 
\frac{1}{2m} \left[ 1 - \frac{ \Gamma(k+1,2m) }{ \Gamma(k+1) } \right],
\label{eq:Pstkg0}
\end{equation}

\noindent
where $\Gamma(s,x)$ is the upper incomplete gamma function,
which can be expressed in terms of the lower incomplete
gamma function, in the form $\Gamma(s,x)= \Gamma(s) - \gamma(s,x)$.
The steady state degree distribution for the
special case of balanced growth and contraction was calculated in Ref. \cite{Moore2006}
for a slightly different model.
The degree distribution $P_{\rm st}(k; \eta=0)$, given by Eq. (\ref{eq:Pstkg0}),
resembles the degree distribution presented in Eq. (20) of Ref. \cite{Moore2006}.
The difference in the pre-factors reflects the variation in the details of the
growth mechanism between the two models.

The discontinuity in the derivative $d w_t/dt |_{t_{\rm vanish}}$ across $\eta= - 1/3$
has interesting implications on the evolution of the degree distribution $P_t(k)$
in the late stages of the contraction process.
For $\eta > - 1/3$ there is a significant time window in which $w_t$ is small and thus
the time dependent degree distribution $P_t(k)$ is in the vicinity of $P_{\rm st}(k)$.
In contrast, for $\eta < - 1/3$ the weight $w_t$ decreases slowly until the very late stages
of the contraction process and then falls down sharply as the time $t_{\rm vanish}$
is approached.
Therefore, there is only an extremely short time window in which
$P_t(k)$ is in the vicinity of $P_{\rm st}(k)$.

As discussed in Sec. IV, the case of $\eta=1$ corresponds to a singular point
of the equation for the generating function $G_t(u)$ [Eq. (\ref{eq:diffeq0a})].
Therefore, this case requires a special treatment.
In Appendix B we solve the master equation for the special case of
pure growth ($\eta=1$) and obtain the time dependent degree
distribution $P_t(k)$ in this case too.
It is given by

\begin{eqnarray}
P_t(k;\eta=1) &=&
\beta_t^{2m+1} P_0(k)
+ \sum_{i=1}^{k} \frac{\beta_t^{2m+1}}{i!} (-2m \ln \beta_t)^i
\left[ P_0(k-i) - P_{\rm st}(k-i; \eta=1) \right]
\nonumber \\
&+& \left( 1 - \beta_t^{2m+1} \right) P_{\rm st}(k; \eta=1),
\label{eq:Ptke1mt}
\end{eqnarray}

\noindent
where $\beta_t$ is given by Eq. (\ref{eq:a_t}) and

\begin{equation}
P_{\rm st}(k;\eta=1) = \frac{1}{2m+1} \left( \frac{2m}{2m+1} \right)^k
\label{eq:Pketa1mt}
\end{equation}

\noindent
is the steady state degree distribution obtained at long times.
Comparing Eq. (\ref{eq:Pktail}) to Eq. (\ref{eq:Pketa1mt}) describing the degree
distribution in the case of pure growth, we conclude that there is a phase
transition at $\eta=1$. 
In the case of pure growth ($\eta=1$) the degree distribution follows an exponential
distribution, whose tail decays more slowly than Eq. (\ref{eq:Pktail})
that applies in the range of $-1 < \eta < 1$.

Consider the special case in which the initial network is generated using the
random attachment model. 
This model is obtained by choosing $\eta=1$, where the number
of edges added in each growth step is denoted by $m_0$ until the
network size reaches $N_0$ nodes.
Using the results of Appendix B, it is found that
for a sufficiently large network size $N_0$ the
generating function of the resulting network
converges towards its steady state form, which
is given by

\begin{equation}
G_0(u) = \frac{1}{2m_0(1-u) + 1}.
\label{eq:G0u}
\end{equation}

\noindent
The initial network is then exposed to a combination of node addition
with random attachment and random node deletion, characterized by
$-1 < \eta < 1$, where the number of edges added in each growth
step is $m$.
Inserting $G_0(u)$ from Eq. (\ref{eq:G0u}) into Eq. (\ref{eq:Ptk0})
and carrying out the differentiation, we obtain

\begin{eqnarray}
P_t(k) &=& \alpha_t^r \frac{ e^{-2 r m(1-\alpha_t) }}{2 m_0 \alpha_t + 1}
\sum_{i=0}^k    
\left( \frac{ 2 m_0 \alpha_t }{ 2 m_0 \alpha_t +1 } \right)^{i}
\frac{ \left[ 2 r m(1-\alpha_t) \right]^{ k-i }  }{ (k-i)! }
\nonumber \\
&+& 
r e^{-2rm} \frac{(2rm)^k}{k!} \int_{\alpha_t}^{1}
y^{r-1} e^{2rmy} (1-y)^k dy.
\label{eq:Ptketain1}
\end{eqnarray}

\noindent
Interestingly, the sum in Eq. (\ref{eq:Ptketain1}) takes the form
of a convolution between an exponential distribution and a 
Poisson distribution.
The mean of the exponential distribution is equal to
$2 m_0 \alpha_t$,
while the mean of the Poisson distribution is
$2rm(1-\alpha_t)$.
The exponential distribution descends from the intial degree distribution,
which is given by Eq. (\ref{eq:Pketa1mt}), 
while the Poisson distribution emerges from the dynamics
of the attachment and deletion processes.
The Poisson distribution describes the degree distribution of an
Erd{\H o}s-R\'enyi network, which is a maximal entropy network
with a given value of the mean degree.
Therefore, the Poisson distribution 
in Eq. (\ref{eq:Ptketain1})
reflects the randomization of the degrees
as the network evolves in time.

Consider the case in which the initial network is an Erd{\H o}s-R\'enyi
network with mean degree $c$, whose degree distribution is known
to be a Poisson distribution.
In this case the time-dependent degree distribution takes a particularly 
simple form, namely

\begin{eqnarray}
P_t(k) &=& \alpha_t^r   e^{ - [ \alpha_t c + 2 r m(1-\alpha_t) ] }
\frac{ \left[ \alpha_t c + 2 r m(1-\alpha_t) \right]^{ k }  }{ k! }
\nonumber \\
&+&
r e^{-2rm} \frac{(2rm)^k}{k!} \int_{\alpha_t}^{1}
y^{r-1} e^{2rmy} (1-y)^k dy.
\label{eq:Ptketain1ER}
\end{eqnarray}

\noindent
The first term in Eq. (\ref{eq:Ptketain1ER}) represents a Poisson distribution
whose mean degree evolves in time, extrapolating between the initial
value of the mean degree, $c$, and a final value of $2rm$.
The second term does not depend on the initial network and is
identical to the corresponding term that is obtained for other 
initial conditions.
In this case the initial network is a maximal entropy network.
For overall network contraction,
under conditions of sufficiently high deletion rate
($-1 < \eta < -1/3$)
the first term of Eq. (\ref{eq:Ptketain1ER}) maintains this property
for a long time window with a decreasing mean degree.
This resembles the behavior in the limit of pure network contraction,
discussed in Refs.
\cite{Tishby2019,Tishby2020}.
 
\begin{figure}
\begin{center}
\includegraphics[width=6cm]{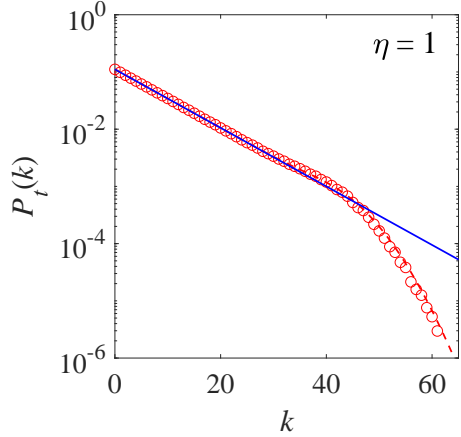} 
\end{center}
\caption{
(Color online)
Analytical results (solid line) for the asymptotic
degree distribution $P_{\rm st}(k)$ of networks that evolves under 
conditions of pure growth ($\eta=1$) via node addition and random attachment
with $m=4$, obtained from Eq. (\ref{eq:Pketa1mt}).
To examine the convergence towards the steady state, 
we also present simulation results (circles)
for the time dependent degree distribution
$P_t(k)$ for a network grown from an initial
ER network of size $N_0=100$ with mean degree $c=3$ up to a size of
$N=10^4$.
The tail of the degree distribution obtained from the simulations 
deviates from the steady state distribution.
This deviation is due to the slow convergence of $P_t(k)$ towards
$P_{\rm st}(k)$ in the case $\eta=1$.
This conclusion is supported by the 
very good agreement between the
simulation results (circles) and the corresponding analytical results (dashed line)
for $P_t(k)$ at $t=N-N_0$, obtained from Eq. (\ref{eq:Ptke1mt}).
}
\label{fig:3}
\end{figure}

In Fig. \ref{fig:3} we present analytical results (solid line),
obtained from Eq. (\ref{eq:Pketa1mt}),
for the steady-state
degree distribution $P_{\rm st}(k)$ of networks that evolve under 
conditions of pure growth ($\eta=1$) via node addition and random attachment
with $m=4$.
To examine the convergence towards the steady-state degree distribution, 
we also present simulation results (circles)
for the time-dependent degree distribution
$P_t(k)$ for a network grown from an initial
ER network of size $N_0=100$ with mean degree $c=3$ up to a size of
$N=10^4$.
The tail of the degree distribution obtained from the simulations 
deviates from the steady state distribution.
This deviation is due to the slow convergence of $P_t(k)$ towards
$P_{\rm st}(k)$ in the case $\eta=1$.
This conclusion is supported by the 
very good agreement between the
simulation results (circles) and the corresponding analytical results
(dashed line) for $P_t(k)$ at $t=N-N_0$, 
obtained from Eq. (\ref{eq:Ptke1mt}).

\begin{figure}
\begin{center}
\includegraphics[width=4.4cm]{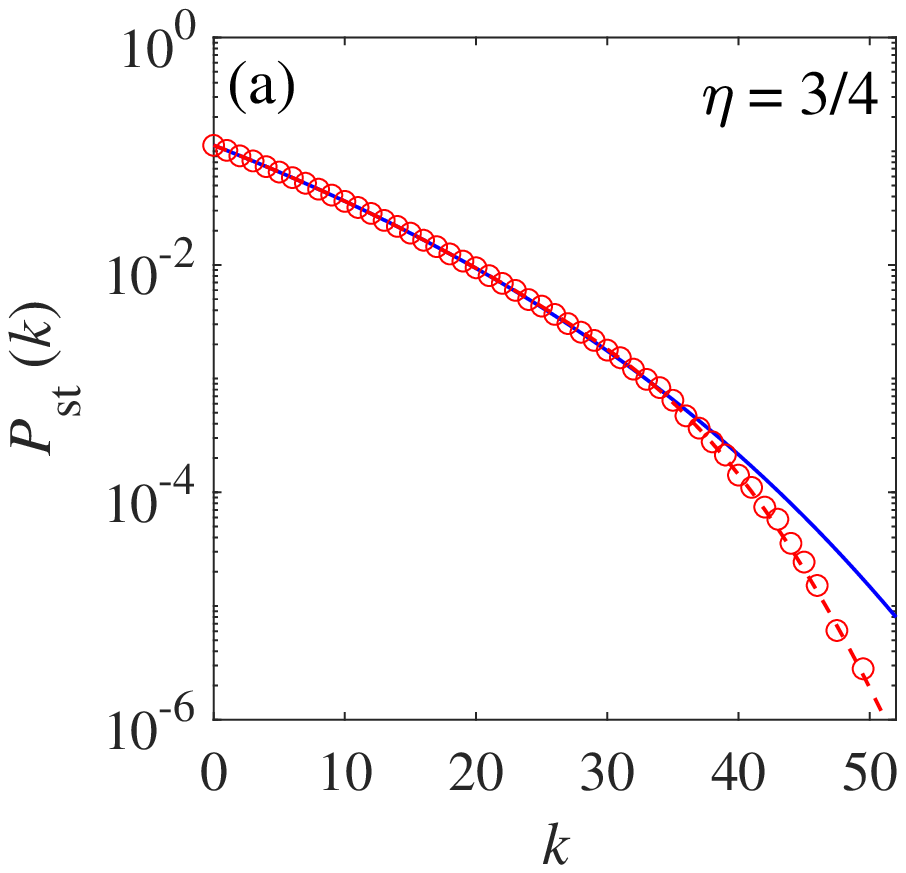} 
\\
\includegraphics[width=4.4cm]{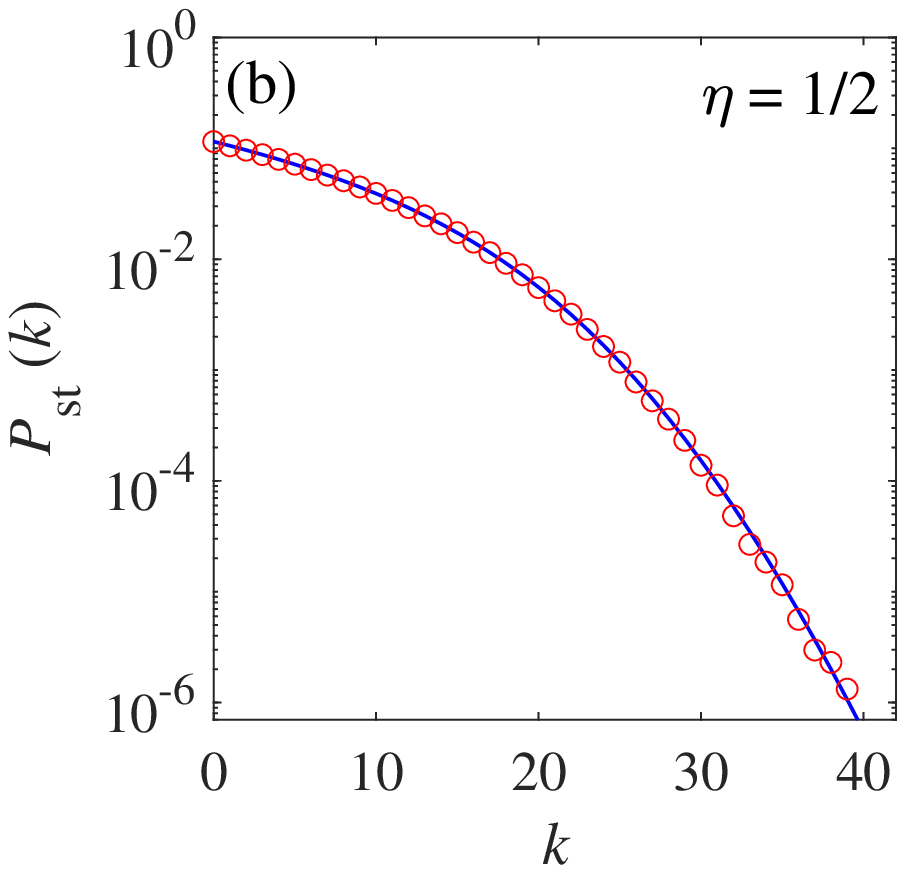} 
\\
\includegraphics[width=4.4cm]{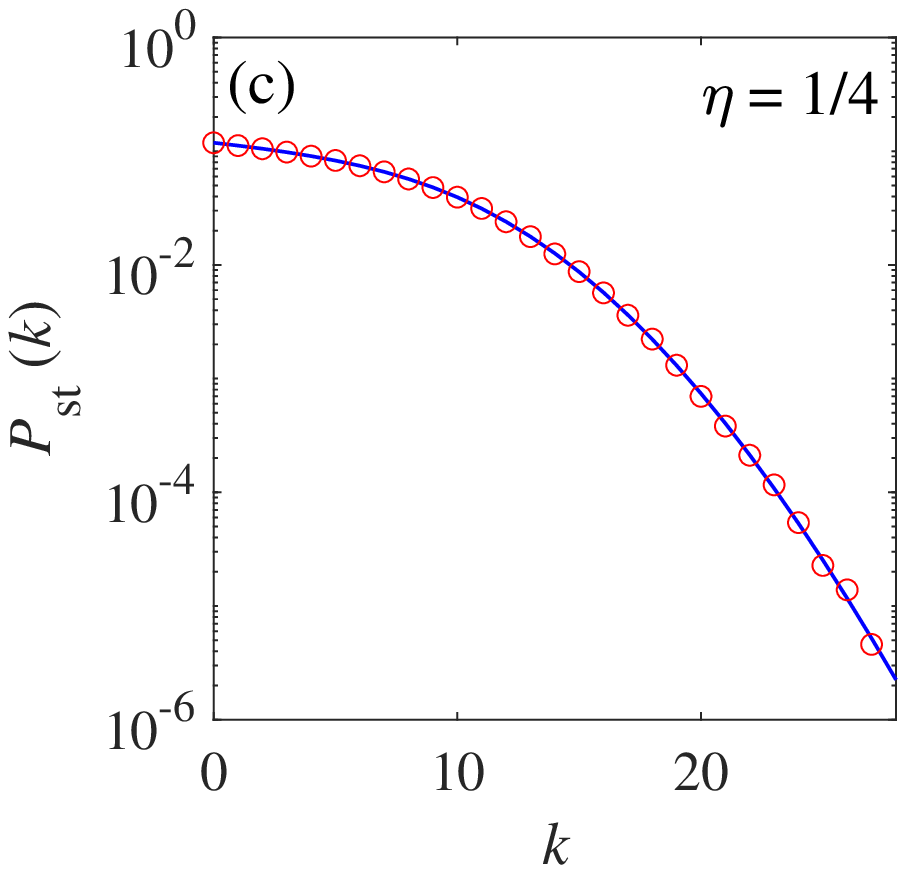} 
\end{center}
\caption{
(Color online)
Analytical results (solid lines), 
obtained from Eq. (\ref{eq:Pstkg}),
for the steady-state
degree distributions $P_{\rm st}(k)$ of networks that evolve under a combination
of growth (via node addition and random attachment) and
contraction (via random node deletion)
in the regime of overall network growth ($0 < \eta < 1$).
Results are presented  
for 
(a) $\eta=3/4$, 
(b) $\eta=1/2$ 
and
(c) $\eta=1/4$.
We also present simulation results (circles), which are shown for $N=10,000$.
The initial network used in the simulations 
is an ER network of size $N_0=100$ with mean degree $c=3$.
In the case of $\eta=1/2$ and $\eta=1/4$ the
analytical results are in very good agreement with the simulation results,
which means that the degree distribution in the simulation
has already converged to its steady-state form
$P_{\rm st}(k)$.
In the case of $\eta=3/4$ one finds that at $N=10,000$ the tail of the
degree distribution $P_t(k)$ deviates from the steady-state distribution 
$P_{\rm st}(k)$. 
This deviation is due to the slow convergence of $P_t(k)$ as $\eta$ is
increased towards $1$.
To justify this conclusion, we also present analytical results (dashed line) for $P_t(k)$,
obtained from Eq. (\ref{eq:Ptketain1ER}) at $t=(N-N_0)/\eta$, which are in very good 
agreement with the simulation results (circles).
}
\label{fig:4}
\end{figure}

In Fig. \ref{fig:4} we present analytical results (solid lines),  
obtained from Eq. (\ref{eq:Pstkg}),
for the steady-state degree distributions $P_{\rm st}(k)$
of networks that evolve under a combination
of growth (via node addition and random attachment) and
contraction (via random node deletion)
in the regime of overall network growth ($0 < \eta < 1$).
Results are presented  
for 
(a) $\eta=3/4$, 
(b) $\eta=1/2$ 
and
(c) $\eta=1/4$.
We also present simulation results (circles), which are shown for $N=10,000$.
The initial network used in the simulations 
is an ER network of size $N_0=100$ with mean degree $c=3$.
In the case of $\eta=1/2$ and $\eta=1/4$ the
analytical results are in very good agreement with the simulation results,
which means that the degree distribution in the simulation
has already converged to its steady-state form
$P_{\rm st}(k)$.
In the case of $\eta=3/4$ one finds that at $N=10,000$ the tail of the
degree distribution $P_t(k)$ deviates from the steady-state distribution 
$P_{\rm st}(k)$. 
This deviation is due to the slow convergence of $P_t(k)$ as $\eta$ is
increased towards $1$.
To justify this conclusion, we also present analytical results (dashed line) for $P_t(k)$,
obtained from Eq. (\ref{eq:Ptketain1ER}) at $t=(N-N_0)/\eta$, which are in very good 
agreement with the simulation results (circles).

\begin{figure}
\begin{center}
\includegraphics[width=5.5cm]{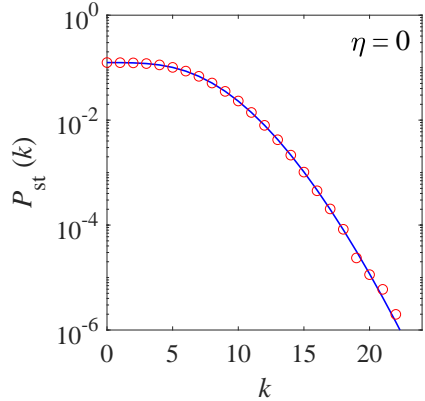} 
\end{center}
\caption{
(Color online)
Analytical results (solid lines) for the asymptotic
degree distributions $P_{\rm st}(k)$ of networks that evolve under a combination
of growth (via node addition and random attachment) and
contraction (via random node deletion)
in the special case of $\eta=0$ in which the network size is fixed,
apart from possible fluctuations.
The initial network is an ER network of size $N_0=10^4$ with mean degree $c=3$.
The analytical results for $P_{\rm st}(k)$ 
are obtained from Eq. (\ref{eq:Pstkg0}). 
The analytical results are in very good agreement with the
simulation results (circles), which are shown for $t=6 N_0$, where
the degree distribution has already converged to its asymptotic form
$P_{\rm st}(k)$.
}
\label{fig:5}
\end{figure}

In Fig. \ref{fig:5} we present analytical results (solid lines),  
obtained from Eq. (\ref{eq:Pstkg0}),
for the steady-state degree distribution $P_{\rm st}(k)$
of networks that evolve under a combination
of growth (via node addition and random attachment) and
contraction (via random node deletion),
in the special case of $\eta=0$ in which 
the network size is fixed,
apart from possible fluctuations.
We also present simulation results (circles).
The initial network is an ER network of size $N_0=10^4$ with mean degree $c=3$.
The analytical results are in very good agreement with the
simulation results (circles), which are shown for $t=6 N_0$, where
the degree distribution has already converged to its asymptotic form
$P_{\rm st}(k)$.

\begin{figure}
\begin{center}
\includegraphics[width=4.2cm]{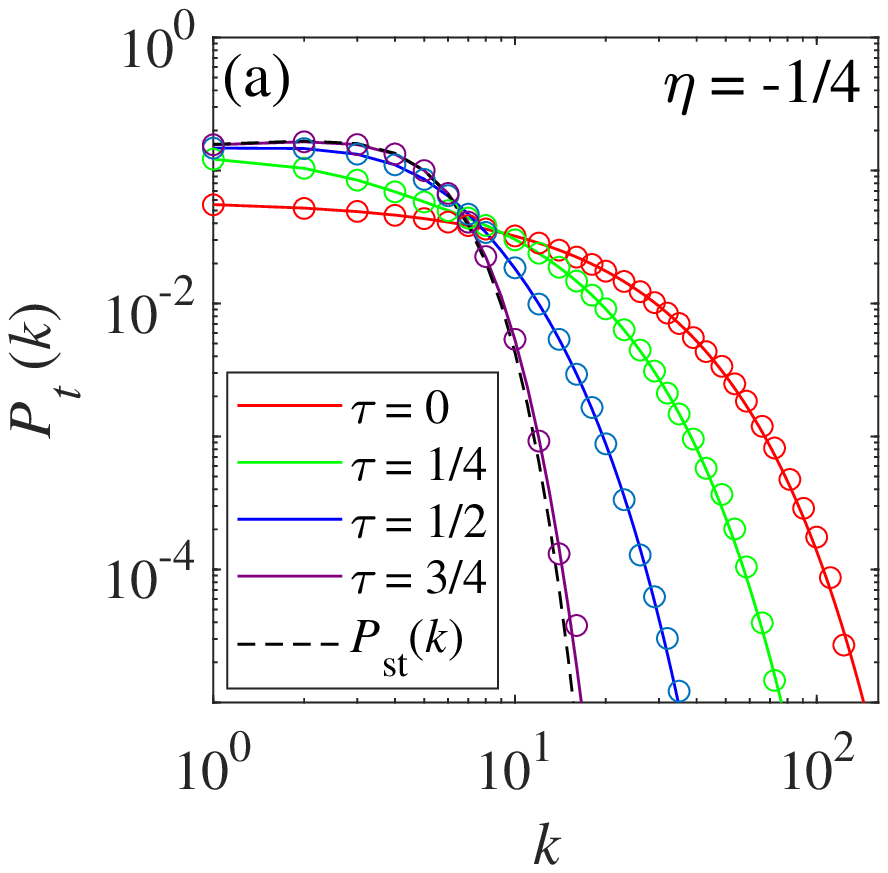}
\\
\includegraphics[width=4.2cm]{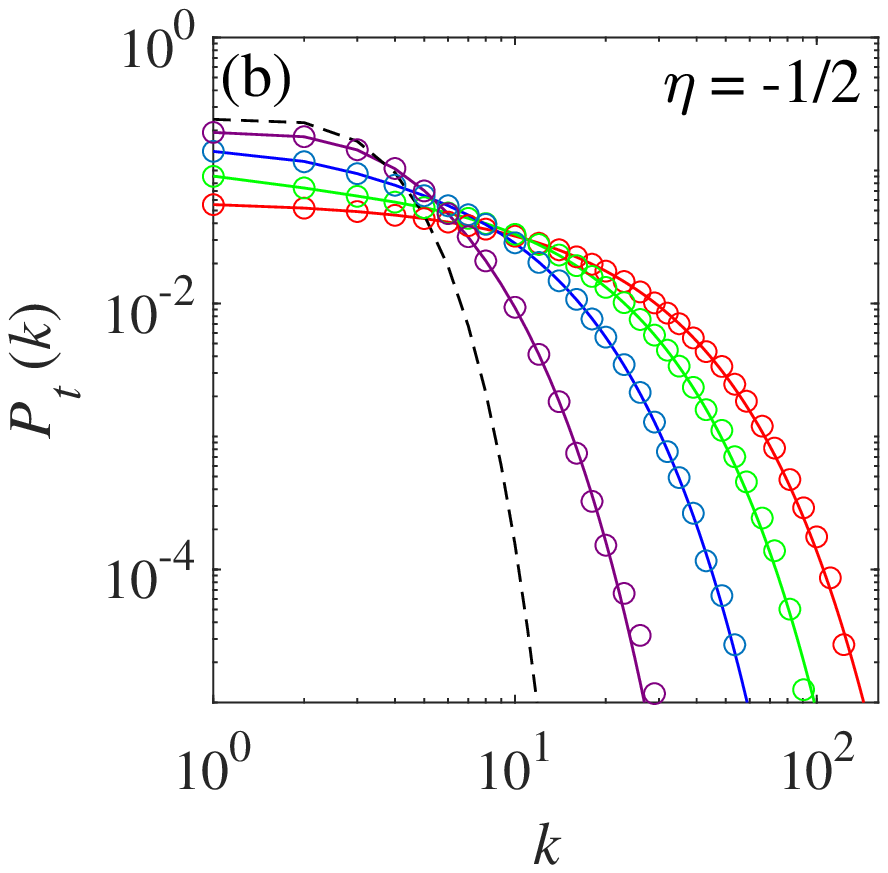}
\\
\includegraphics[width=4.2cm]{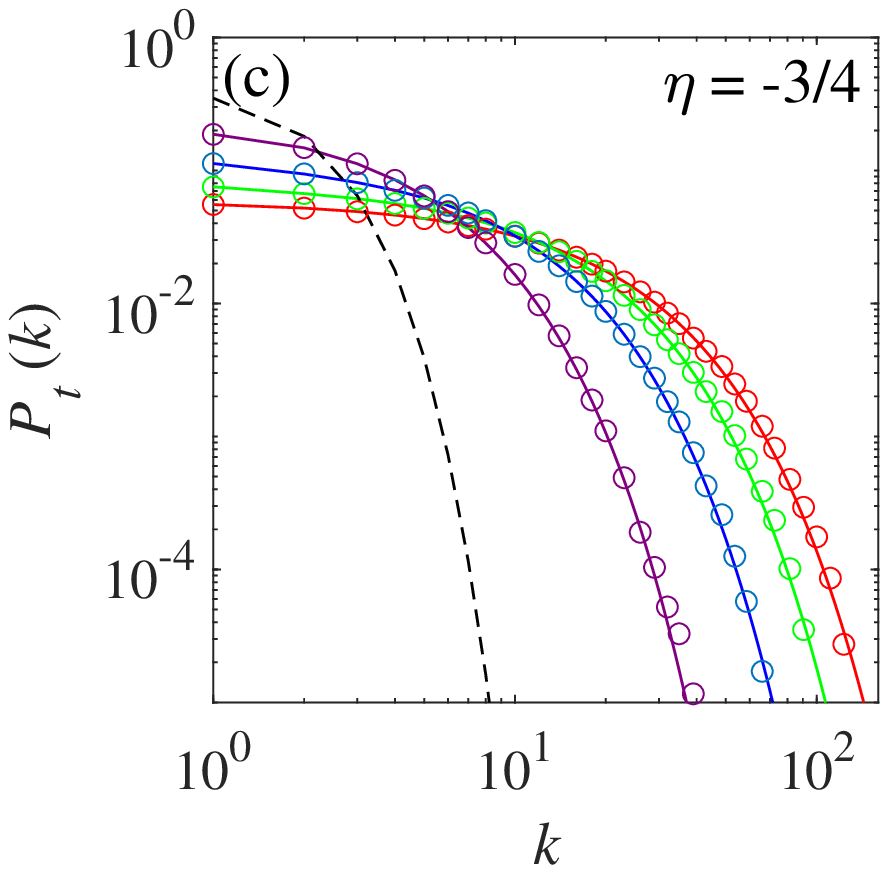}
\end{center}
\caption{
(Color online)
Analytical results (solid lines) for the
degree distributions of networks that evolve under a combination
of growth via random node addition and random attachment and
contraction via random node deletion in the regime of overall network contraction
for (a) $\eta=-1/4$, (b) $\eta=-1/2$ and (c) $\eta=-3/4$.
In each frame the degree distribution $P_t(k)$ is shown
(right to left) for
$\tau=0$,
$\tau=1/4$,
$\tau=1/2$
and $\tau=3/4$,
where the normalized time $\tau$ is the fraction 
of nodes that have been deleted [Eq. (\ref{eq:tau})].
The asymptotic distribution $P_{\rm st}(k)$ is also shown
(dashed lines).
The initial network is obtained from random node addition and random attachment
with $m_0=8$ and it consists of $N_0=12,500$ nodes.
The analytical results for $P_t(k)$,
are obtained from Eq. (\ref{eq:Ptketain1}).
The simulation results (circles) are in very good agreement with the 
corresponding analytical results.
As time evolves the time dependent degree distribution $P_t(k)$ converges towards
the asymptotic distribution $P_{\rm st}(k)$.
For $\eta=-1/4$, the degree distribution $P_t(k)$ approaches 
$P_{\rm st}(k)$ when
a significant fraction of the network is still in place.
In contrast, for $\eta=-1/2$ and $-3/4$ 
the convergence of $P_t(k)$ is initially very slow and it gets
closer to $P_{\rm st}(k)$ only shortly before the network vanishes.
The transition between the two dynamical behaviors
takes place at $\eta=-1/3$.
}
\label{fig:6}
\end{figure}

In Fig. \ref{fig:6} we present analytical
results (solid lines) for the time-dependent degree distributions $P_t(k)$
of networks that evolve under a combination
of growth (via node addition and random attachment) and
contraction (via random node deletion) 
in the regime of overall network contraction
for (a) $\eta=-1/4$, (b) $\eta=-1/2$ and (c) $\eta=-3/4$.
In each frame the degree distribution $P_t(k)$,
obtained from Eq. (\ref{eq:Ptketain1}),
is shown (right to left) for
$\tau=0$,
$\tau=1/4$,
$\tau=1/2$
and $\tau=3/4$, 
where the normalized time $\tau$ is the fraction 
of nodes that have been deleted [Eq. (\ref{eq:tau})].
The long-time degree distribution $P_{\rm st}(k)$,
obtained from Eq. (\ref{eq:Pstkg}),
is also shown
(dashed lines).
The initial condition at $t=0$
is a network obtained from random node addition and random attachment
with $m_0=8$ and it consists of $N=12,500$ nodes.
Thus, the initial degree distribution $P_0(k)$ is given by Eq. (\ref{eq:Pketa1mt}),
with $m$ replaced by $m_0$.
The simulation results (circles) are in very good agreement with the 
corresponding analytical results.
As time evolves the time dependent degree distribution $P_t(k)$ converges towards
the asymptotic distribution $P_{\rm st}(k)$.
For $\eta=-1/4$ the degree distribution $P_t(k)$ approaches
$P_{\rm st}(k)$ when
a significant fraction of the network is still in place.
In contrast, for $\eta=-1/2$ and $-3/4$ 
the convergence of $P_t(k)$ is initially very slow and it gets
closer to $P_{\rm st}(k)$ only shortly before the network vanishes.
The transition between the two dynamical behaviors
takes place at $\eta=-1/3$.

\section{The mean and variance of the degree distribution}

The mean degree at time $t$ can be obtained from

\begin{equation}
\langle K \rangle_t = \frac{d}{du} G_t(u) \bigg\vert_{u=1}.
\label{eq:<K>t}
\end{equation}

\noindent
Inserting $G_t(u)$ from Eq. (\ref{eq:Gtu20}) into Eq. (\ref{eq:<K>t}),
we obtain

\begin{equation}
\langle K \rangle_t = \alpha_t^{r+1} \langle K \rangle_0
+ (1-\alpha_t^{r+1}) \langle K \rangle_{\rm st},
\label{eq:<K>t2}
\end{equation}

\noindent
where

\begin{equation}
\langle K \rangle_{\rm st} = \frac{2rm}{r+1}.
\end{equation}

\begin{figure}
\begin{center}
\includegraphics[width=5.5cm]{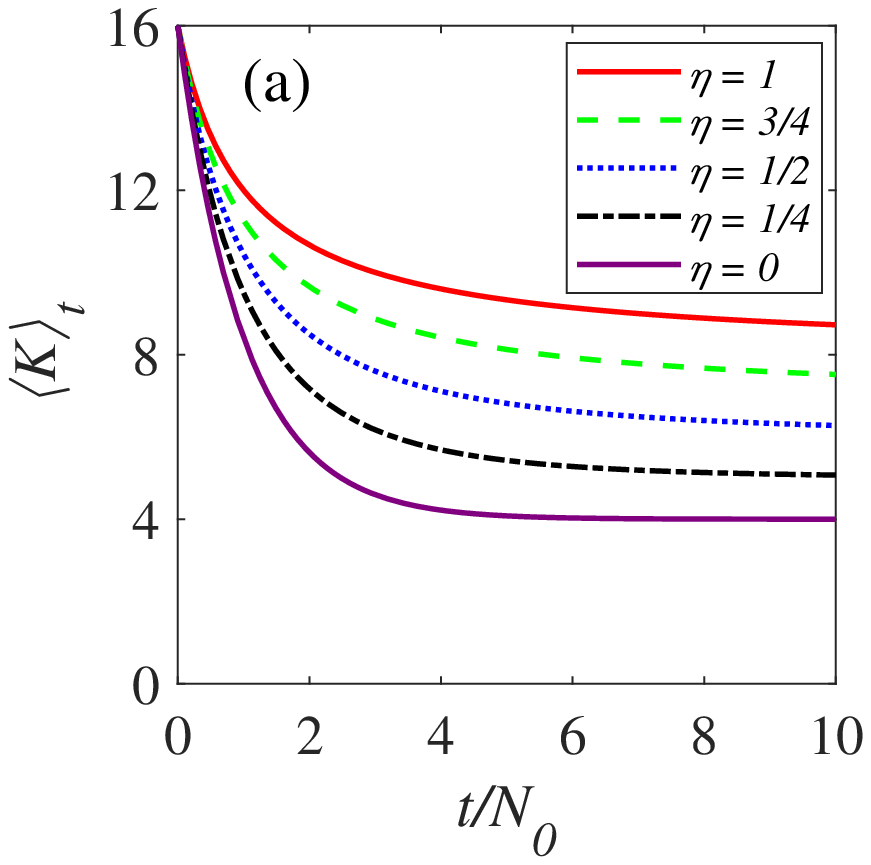}
\\
\includegraphics[width=5.5cm]{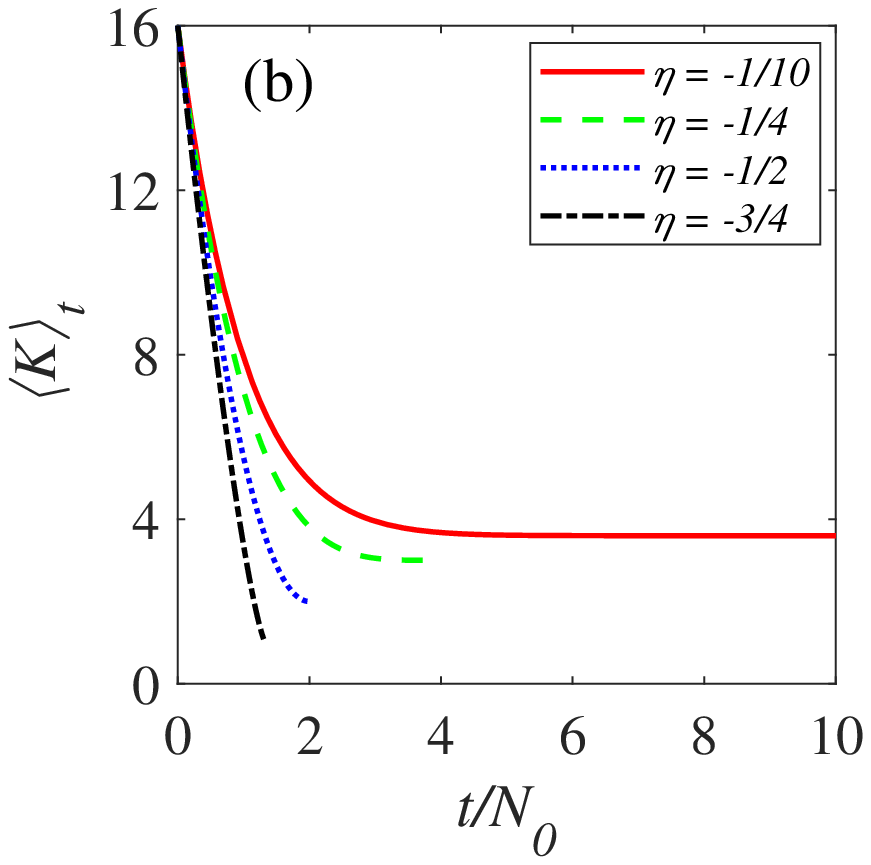}
\end{center}
\caption{
(Color online)
Analytical results (solid lines),
obtained from Eq. (\ref{eq:<K>t2}),
for the
mean degree $\langle K \rangle_t$ vs. time $t$
for networks that evolve under a combination
of growth (via node addition and random attachment) and
contraction (via random node deletion)
for (a) $\eta=1$, $3/4$, $1/2$, $1/4$ and $0$ (from top to bottom);
and (b) $\eta=-1/10$, $-1/4$, $-1/2$ and $-3/4$ (from top to bottom).
In all cases the initial network has a mean degree of $\langle K \rangle_0=16$.
In case that $\eta>0$ the mean degree gradually converges towards
its asymptotic value. 
In case that $\eta<0$ the network vanishes at a finite time $t_{\rm vanish}=N_0/|\eta|$.
}
\label{fig:7}
\end{figure}

In Fig. \ref{fig:7} we present analytical
results (solid lines),
obtained from Eq. (\ref{eq:<K>t2}),
for the
mean degree $\langle K \rangle_t$ vs. time $t$
for networks that evolve under a combination
of growth (via node addition and random attachment) and
contraction (via random node deletion) 
for (a) $0 \le \eta < 1$; and (b) $-1 < \eta < 0$.
The mean degree of the initial network is $\langle K \rangle_0 = 16$.
In case that $\eta>0$ the mean degree gradually converges towards
its asymptotic value. 
In case that $\eta<0$ the network vanishes at a finite time $t_{\rm vanish}=N_0/|\eta|$.

\begin{figure}
\begin{center}
\includegraphics[width=5.5cm]{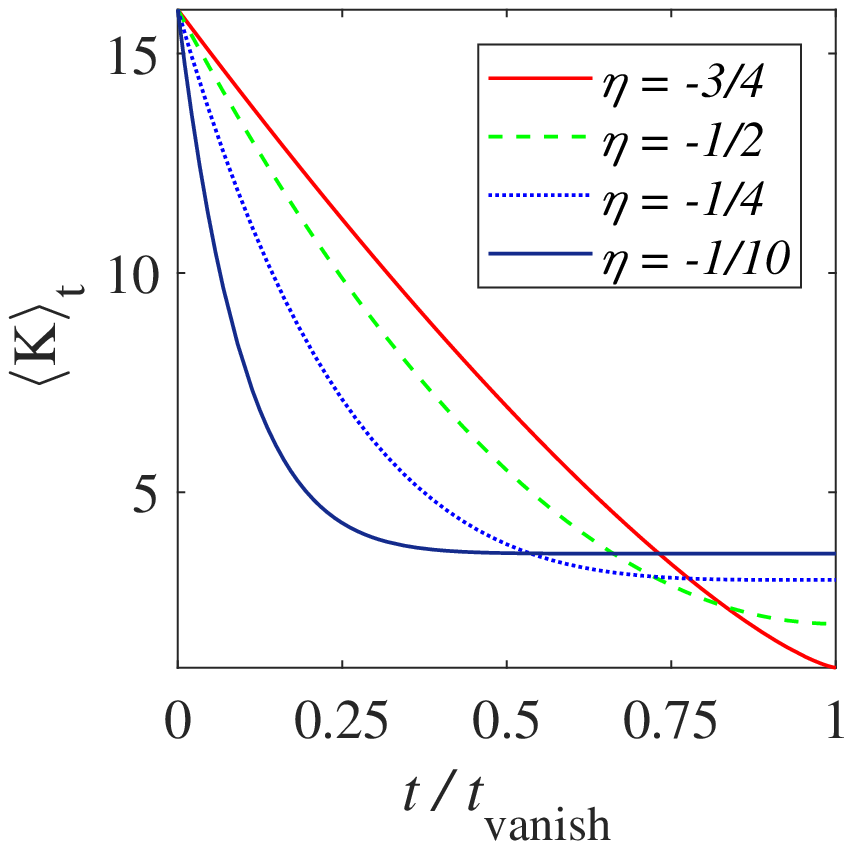}
\end{center}
\caption{
(Color online)
Analytical results (solid lines),
obtained from Eq. (\ref{eq:<K>t2}),
for the
mean degree $\langle K \rangle_t$ vs. $t/t_{\rm vanish}$
for networks that evolve under a combination
of growth (via node addition and random attachment) and
contraction (via random node deletion)
for $\eta=-3/4$, $-1/2$, $-1/4$ and $-1/10$ (from top to bottom).
The initial network has a mean degree of $\langle K \rangle_0=16$.
}
\label{fig:8}
\end{figure}

In Fig. \ref{fig:8} we present analytical
results (solid lines),
for the
mean degree $\langle K \rangle_t$ vs. $t/t_{\rm vanish}$
for networks that evolve under a combination
of growth (via node addition and random attachment) and
contraction (via random node deletion) 
for $-1 < \eta < 0$.

To obtain the variance ${\rm Var}_t(K)$ we use the cumulant generating
function, which is given by

\begin{equation}
F_t(x) = \ln G_t(e^x).
\label{eq:Ftx}
\end{equation}

\noindent
The variance is obtained from

\begin{equation}
{\rm Var}_t(K) = 
\frac{d^2}{dx^2} F_t(x) \bigg\vert_{x=0}.
\label{eq:vark}
\end{equation}

\noindent
Inserting $F_t(x)$ from Eq. (\ref{eq:Ftx}) into Eq. (\ref{eq:vark})
we obtain

\begin{eqnarray}
{\rm Var}_t(K) &=& 
\alpha_t^{r+2} {\rm Var}_0(K)
+ \alpha_t^{r+1} \left[ (\alpha_t-1) \langle K \rangle_0^2 
+ (\alpha_t^{r+1} - 2 \alpha_t + 1) \langle K \rangle_0 \right]
\nonumber \\
&-& \alpha_t^{r+1} (\alpha_t^{r+1} - 1) 
\left( \langle K \rangle_0 - \langle K \rangle_{\rm st} \right)^2
\nonumber \\
&+& 2 \alpha_t^{r+1} (\alpha_t -1) (r+1) 
\left[ \frac{r+1}{r+2} \langle K \rangle_{\rm st} - \langle K \rangle_{0} \right] 
\langle K \rangle_{\rm st}
\nonumber \\
&+& (1-\alpha_t^{r+1}) {\rm Var}_{\rm st}(K),
\label{eq:VartK}
\end{eqnarray}

\noindent
where

\begin{equation}
{\rm Var}_{\rm st}(K) = 
\frac{2rm[(2m+1) r^2 + 3 r + 2 ]}{(r+1)^2 (r+2)} 
\end{equation}

\noindent
is the variance of $P_{\rm st}(k)$, given by Eq. (\ref{eq:Pstkg}).
Note that at $t=0$ the right hand side of Eq. (\ref{eq:VartK}) is reduced
to ${\rm Var}_0(K)$ while in the long time limit it converges towards
${\rm Var}_{\rm st}(K)$.

The mean $\langle K \rangle_t(\eta=1)$ and variance 
${\rm Var}_t(K;\eta=1)$
of the degree distribution
$P_t(k;\eta=1)$ in the case of $\eta=1$ are
calculated in Appendix B.
The steady state results $\langle K \rangle_{\rm st}(\eta=1)$
and ${\rm Var}_{\rm st}(K;\eta=1)$ coincide with those obtained
from $\langle K \rangle_t$ and ${\rm Var}_t(K)$, respectively,
in the limit of $\eta \rightarrow 1$ ($r \rightarrow \infty$).

\section{Summary and Discussion}

We presented analytical results for the time-dependent degree distribution
$P_t(k)$ of networks that evolve under the combination of
growth (via node addition and random attachment) 
and contraction (via random node deletion).
In case that the rate of node addition exceeds the rate of node deletion,
the overall process is of network growth, while in the opposite case
the overall process is of network contraction.
Using the master equation and the generating function formalism  
we obtained a closed form expression for the degree distribution $P_t(k)$.
It includes a term that depends on the initial condition $P_0(k)$, 
which decays as time evolves,
and a long-time asymptotic term $P_{\rm st}(k)$,
which is an attractive fixed point.
Interestingly, the expression for $P_t(k)$ is identical in the 
regimes of overall growth and overall contraction.

The model of network growth via node addition and random attachment
can be considered as the simplest network growth model.
It gives rise to networks that exhibit an exponential degree distribution.
Similarly, the model of network contraction via random node deletion
can be considered as the simplest network contraction model.
The contracting networks converge towards the ER structure, which
exhibits a Poisson degree distribution whose mean degree decreases
as time proceeds.
The combination of growth via node addition and random attachment
and contraction via random node deletion yields novel structures which
depend on the balance between the rates of the two processes. 

\begin{figure}
\begin{center}
\includegraphics[width=8cm]{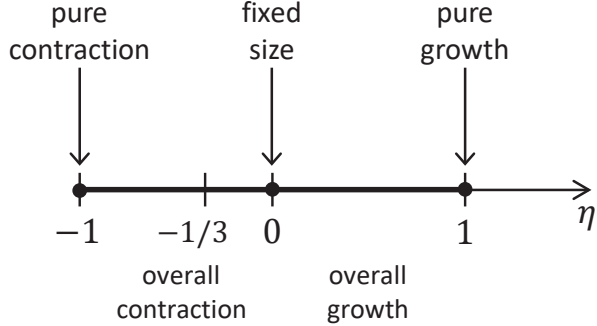}
\end{center}
\caption{
The phase diagram of networks that evolve under a combination
of growth via random node addition and random attachment 
and contraction via random node deletion, in terms of the
growth rate $-1 \le \eta \le 1$.
The case of $\eta=1$ represents pure network growth
via node addition and random attachment.
The case of $0 < \eta < 1$ represents a combination of growth and
contraction where the overall process is of network growth.
The case of $\eta=0$ represents a balance between the growth and
contraction processes such that on average the network size remains fixed.
The case of $-1 < \eta < 0$ represents a combination of growth and
contraction where the overall process is of network contraction.
The case of $\eta=-1$ corresponds to pure contraction via 
random node deletion.
At $\eta=1$ there is a structural phase transition between the exponential
degree distribution in the asymptotic state for $\eta=1$ and the 
asymptotic Poisson-like degree distribution
in the regime of $0 < \eta <1$, whose tail decays faster than
the exponential distribution.
At $\eta=0$ there is a phase transition between the $\eta>0$
phase which exhibits an ever growing network whose degree
distribution converges to an asymptotic form and
the $\eta<0$ phase in which the network vanishes after a finite time $t_{\rm vanish}$.
At $\eta=-1/3$ there is a dynamical transition.
For $-1/3 < \eta < 0$ the degree distribution $P_t(k)$ quickly converges towards
$P_{\rm st}(k)$.
In contrast, for $-1 < \eta < -1/3$ 
the convergence of $P_t(k)$ is initially very slow and it gets
closer to $P_{\rm st}(k)$ only shortly before the network vanishes.
}
\label{fig:9}
\end{figure}

In Fig. \ref{fig:9} we present the
phase diagram of networks that evolve under a combination
of growth (via node addition and random attachment) 
and contraction (via random node deletion), in terms of the
growth rate $-1 \le \eta \le 1$.
The case of $\eta=1$ represents pure network growth
via node addition and random attachment.
The case of $0 < \eta < 1$ represents a combination of growth and
contraction where the overall process is of network growth.
The case of $\eta=0$ represents a balance between the growth and
contraction processes such that on average the network size remains fixed.
The case of $-1 < \eta < 0$ represents a combination of growth and
contraction where the overall process is of network contraction.
The case of $\eta=-1$ corresponds to pure contraction via 
random node deletion.

At $\eta=1$ there is a structural phase transition between the steady-state degree
distribution at $\eta=1$, which follows an exponential distribution,
given by Eq. (\ref{eq:Pketa1mt}),
and the steady-state degree distribution
in the regime of $0 < \eta <1$,
given by Eq. (\ref{eq:Pstkg}),
which decays like a Poisson distribution.
This degree distribution essentially consists of a linear combination
of Poisson distributions. 
Its tail is dominated by the Poisson component with the largest mean
degree, given by Eq. (\ref{eq:Pktail}).
This transition implies that even the slightest rate of node deletion
leads to a qualitative change in the nature of the 
steady state degree distribution.
From a technical point of view, $\eta=1$ is a singular point in the
differential equation (\ref{eq:diffeq0a}) for the generating function $G_t(u)$,
where the order of the equation changes.
The phase transition at $\eta=1$ essentially emanates from this singularity.

At $\eta=0$ there is a phase transition between the $\eta>0$
phase which exhibits an ever growing network 
and the $\eta<0$ phase in which the network vanishes after a finite time.
Surprisingly, the expression for the time dependent degree distribution $P_t(k)$,
given by Eq. (\ref{eq:Ptketain1}),
is identical on both sides of the transition.
However, the qualitative behavior of the coefficient $\alpha_t$
is fundamentally different on both sides. 
For $\eta>0$ the coefficient $\alpha_t$ gradually decays as time evolves
but remains positive at any finite time.
In contrast, for $\eta<0$ it decays to zero after a finite time $t_{\rm vanish}$,
at which the whole network vanishes.

At $\eta=-1/3$ there is a dynamical transition between a phase of slow network 
contraction for $-1/3 < \eta < 0$ and a fast contracting phase for $-1 \le \eta < -1/3$.
In the phase of slow contraction  
the degree distribution converges towards $P_{\rm st}(k)$ 
and remains in its vicinity for a finite time window, before the network vanishes.
In the fast contracting phase the network size quickly decreases
and it vanishes before the weight of $P_{\rm st}(k)$ becomes significant.
In this case, the evolution of the degree distribution $P_t(k)$ during the
contraction process qualitatively resembles the case of pure network contraction
via random node deletion ($\eta=-1$),
considered in Refs. \cite{Tishby2019,Tishby2020}.

The behavior of the degree distribution $P_t(k)$ in the scenario
of overall network contraction $-1 < \eta < 0$ can be considered in the 
context of dynamical processes that exhibit intermediate asymptotic states
\cite{Barenblatt1996,Barenblatt2003}.
These are states that appear at intermediate time scales,
which are sufficiently long for such structures to build up,
but shorter than the time scales at which the whole system
disintegrates. 
The intermediate time scales can be made arbitrarily long by increasing
the initial size of the system, justifying the term `asymptotic'.
More specifically, in the regime of $-1/3 < \eta < 0$ the intermediate
asymptotic state exhibits the degree distribution $P_{\rm st}(k)$,
while in the regime of $-1 \le \eta < -1/3$ the intermediate
asymptotic degree distribution is dominated by the first term of $P_t(k)$,
given by Eq. (\ref{eq:Ptk0}).

This work was supported by grant no. 2020720 from the United States-Israel 
Binational Science Foundation (BSF).

\appendix
\section{Calculation of the degree distribution $P_t(k)$}

In this Appendix we solve the master equation [Eq. (\ref{eq:dP(t)/dtRC0})]
for $-1 \le \eta < 1$ and obtain the time dependent degree distribution $P_t(k)$.
In the first step we 
solve the differential equation (\ref{eq:diffeq0a})
using the method of characteristics
and obtain the time dependent generating
function $G_t(u)$.
The method of characteristics applies to hyperbolic partial differential equations.
In this method the partial differential equation is reduced to a set of ordinary
differential equations called characteristic equations.

The characteristic equations of Eq. (\ref{eq:diffeq0a}) can be written as

\begin{equation}
\frac{du}{dt}  =  - \frac{1-\eta}{2} \frac{1-u}{N_0+\eta t}
\label{eq:C1}
\end{equation}

\noindent
and

\begin{equation}
\frac{d G_t(u)}{du}  =  \frac{1+\eta}{1-\eta} 
\left[ \left( 2m + \frac{1}{1-u} \right) G_t(u) - \frac{1}{1-u} \right].
\label{eq:C2}
\end{equation}

\noindent
Solving Eq. (\ref{eq:C1}), one obtains a relation between $u$ and $t$,
via an integration constant $C_1$.
In the case of $\eta \ne 0$, it is given by

\begin{equation}
C_1 = 
\frac{(1-u)^{ \frac{2 \eta}{1-\eta} } }{N_0+\eta t},    
\end{equation}

\noindent
while in the case of $\eta=0$ it is given by

\begin{equation}
C_1 =
(1-u)  e^{-t/2 N_0}.
\label{eq:C_1eta0}
\end{equation}

\noindent
In order to solve Eq. (\ref{eq:C2}), we express the generating function
in the form

\begin{equation}
G_t(u) = G_t^{(h)}(u) + G_t^{(p)}(u),
\label{eq:GtuA}
\end{equation}

\noindent
where $G_t^{(h)}(u)$ is the homogeneous part and $G_t^{(p)}$ is
the inhomogeneous part of $G_t(u)$.
Solving for the homogeneous part, we obtain

\begin{equation}
G_t^{(h)}(u) = C_2 e^{2 r m u} (1-u)^{-r},
\label{eq:Gh}
\end{equation}

\noindent
where $C_2$ is an integration constant, and $r$ is defined in Eq. (\ref{eq:r}).
Solving Eq. (\ref{eq:C2}) for the inhomogeneous part of $G_t(u)$,
we obtain

\begin{equation}
G_t^{(p)}(u) = r e^{-2rm(1-u)} 
\frac{ \gamma[r,-2rm(1-u)] }{[-2rm(1-u)]^r},
\label{eq:Ginh}
\end{equation}

\noindent
where

\begin{equation}
\gamma(s,x) = \int_{0}^{x} t^{s-1} e^{-t} dt
\label{eq:lowergamma}
\end{equation}

\noindent
is the lower incomplete gamma function
\cite{Olver2010}.
Inserting $G_t^{(h)}(u)$ from Eq. (\ref{eq:Gh}) and 
$G_t^{(p)}(u)$ from Eq. (\ref{eq:Ginh})
into Eq. (\ref{eq:GtuA})
and extracting the integration constant $C_2$, we obtain

\begin{equation}
C_2 = e^{-2rmu} (1-u)^r G_t(u) 
- r e^{-2rm} (1-u)^r 
\frac{ \gamma[r,-2rm(1-u)] }{ (-2rm)^r }.
\end{equation}

\noindent
Starting with the case of $\eta \ne 0$, we combine the solutions of the
two characteristic equations and
obtain the solution of Eq. (\ref{eq:diffeq0a}),
which is given by

\begin{equation}
G_t(u) = e^{2 r m u} (1-u)^{-r} F \left[ \frac{(1-u)^{\frac{2 \eta}{1-\eta}}}{N_0+\eta t} \right]
+ r e^{-2 r m(1-u)} 
\frac{ \gamma[r,-2rm(1-u)] }{ [-2rm(1-u)]^r },
\label{eq:Gtu0}
\end{equation}

\noindent
where $F$ is an arbitrary function.
In order to impose the initial condition $G_0(u)$  
we set $t=0$ in Eq. (\ref{eq:Gtu0}) and obtain

\begin{equation}
G_0(u) = e^{2 r m u} (1-u)^{-r} F \left[ \frac{(1-u)^{\frac{2 \eta}{1-\eta}}}{ N_0 } \right]
+ r e^{-2 r m(1-u)} 
\frac{ \gamma[r,-2rm(1-u)] }{ [-2rm(1-u)]^r }.
\label{eq:Gtu0b}
\end{equation}

\noindent
Solving for the arbitrary function $F$, we obtain

\begin{equation}
F \left[ \frac{(1-u)^{\frac{2 \eta}{1-\eta}}}{ N_0 } \right] =
e^{-2 r m u} (1-u)^{r} G_0(u)    
- r e^{-2 r m } 
\frac{ \gamma[r,-2rm(1-u)] }{ (-2rm)^r }.
\label{eq:Ftu0b}
\end{equation}

We introduce the variable

\begin{equation}
z =
\frac{(1-u)^{\frac{2 \eta}{1-\eta}}}{N_0}.
\end{equation}

\noindent
Expressing $u$ in terms of $z$, we obtain

\begin{equation}
u = 1 - (z N_0)^{ \frac{1-\eta}{2 \eta} }.
\end{equation}

\noindent
Rewriting Eq. (\ref{eq:Ftu0b}) in terms of the variable $z$, we obtain

\begin{eqnarray}
F(z) &=& e^{-2rm \left[ 1 -(zN_0)^{ \frac{1-\eta}{2 \eta} } \right] } 
(z N_0)^{ \frac{ r (1-\eta) }{2 \eta} } G_0 \left[ 1 -(zN_0)^{ \frac{1-\eta}{2 \eta} } \right]
\nonumber \\
&-& r e^{-2 r m}
\frac{ \gamma \left[ r,-2rm (z N_0)^{\frac{1-\eta}{2 \eta} } \right] }{(-2rm)^r}.
\label{eq:Fu}
\end{eqnarray}

\noindent
Inserting $F(z)$ from Eq. (\ref{eq:Fu}) into Eq. (\ref{eq:Gtu0}), we obtain

\begin{eqnarray}
G_t(u) &=& 
\alpha_t^r e^{ -2rm(1-u)(1-\alpha_t) } G_0[1-\alpha_t(1-u)]
\nonumber \\
&+& r e^{-2 r m (1-u)} 
\frac{ \gamma[r,-2rm(1-u)] - \gamma[r,-2rm \alpha_t(1-u)] }{ [-2rm(1-u)]^r },
\label{eq:Gtu1}
\end{eqnarray}

\noindent
where 

\begin{equation}
\alpha_t = \left( 1 + \frac{\eta t}{N_0} \right)^{ - \frac{ 1-\eta }{ 2 \eta } }.
\label{eq:alphatA}
\end{equation}

\noindent
A similar analysis applies to the special case of $\eta=0$.
In this case one needs to use the special expression for $C_1$,
given by Eq. (\ref{eq:C_1eta0}). 
It yields the same form of $G_t(u)$, given by Eq. (\ref{eq:Gtu1}),
but with a different expression for $\alpha_t$, which in the case
of $\eta=0$ is given by 

\begin{equation}
\alpha_t = \exp \left( - \frac{t}{2 N_0} \right).
\label{eq:alphatA0}
\end{equation}

To simplify Eq. (\ref{eq:Gtu1}) we first denote

\begin{equation}
S(u) = \gamma[r,-2rm(1-u)] - \gamma[r,-2rm \alpha_t (1-u)].
\end{equation}

\noindent
Replacing $\gamma(s,x)$ by its integral representation (\ref{eq:lowergamma}),
one can express $S(u)$ in the form

\begin{equation}
S(u) = \int_{-2rm \alpha_t (1-u)}^{-2rm(1-u)} x^{r-1} e^{-x} dx.
\label{eq:Su}
\end{equation}

\noindent
Substituting $x=-2rm(1-u)y$ in Eq. (\ref{eq:Su}), we obtain

\begin{equation}
S(u) = [-2rm(1-u)]^r \int_{\alpha_t}^{1} y^{r-1} e^{2rm(1-u)y} dy.
\label{eq:Su2}
\end{equation}

\noindent
Plugging $S(u)$ from Eq. (\ref{eq:Su2}) into Eq. (\ref{eq:Gtu1}),
one obtains

\begin{eqnarray}
G_t(u) &=& 
\alpha_t^r e^{ -2rm(1-u)(1-\alpha_t) } G_0[1-\alpha_t(1-u)]
\nonumber \\
&+& r 
\int_{\alpha_t}^{1} y^{r-1} e^{-2rm(1-u)(1-y)} dy.
\label{eq:Gtuint}
\end{eqnarray}

The time dependent degree distribution is 
obtained by differentiating the generating function $G_t(u)$:

\begin{equation}
P_t(k) = \frac{1}{k!} \frac{ \partial^k G_t(u) }{ \partial u^k }  \bigg\vert_{u=0}.
\label{eq:PtkD}
\end{equation}
 
\noindent
Inserting $G_t(u)$ from Eq. (\ref{eq:Gtuint}) into Eq. (\ref{eq:PtkD}),
we obtain the main result of this Appendix, namely

\begin{eqnarray}
P_t(k) &=& \alpha_t^r \frac{ e^{-2 r m(1-\alpha_t) }}{k!}
\sum_{i=0}^k \binom{k}{i} \alpha_t^i 
\frac{d^i G_0(u)}{du^i} \bigg\vert_{u=1-\alpha_t}
\left[ 2 r m(1-\alpha_t) \right]^{ k-i }  
\nonumber \\
&+& r e^{-2rm} \frac{(2rm)^k}{k!}  
\int_{\alpha_t}^{1} y^{r-1} e^{2rmy}(1-y)^k dy.
\label{eq:Ptk}
\end{eqnarray}

\noindent
This is a closed form analytical expression for the time dependent degree distribution
$P_t(k)$. It is based on the initial degree distribution $P_0(k)$, which is encoded in the
generating function at time $t=0$, $G_0(u)$.

\section{Calculation of $P_t(k)$ in the case of pure network growth}

The case of pure network growth via node addition and random 
attachment is obtained for $\eta=1$.
Inserting $\eta=1$ in Eq. (\ref{eq:diffeq0a}), we obtain

\begin{equation}
(N_0 +   t) \frac{ \partial G_t(u;\eta=1) }{\partial t} 
=
-   \left[ 2m(1-u) + 1 \right] G_t(u;\eta=1) + 1.
\label{eq:diffeq0}
\end{equation}
 
\noindent
The characteristic equations in this case are given by

\begin{equation}
\frac{d u}{d t} = 0,
\label{eq:dut0}
\end{equation}
 
\noindent
and

\begin{equation}
\frac{d G_t(u;\eta=1)}{dt} = \frac{ 1 - [2m(1-u) + 1]G_t(u;\eta=1) }{N_0 + t}.
\label{eq:dGtu0}
\end{equation}
 
\noindent
From Eq. (\ref{eq:dut0}) one finds that on the characteristic lines
the variable $u$ is a constant that does not depend on time.
Solving Eq. (\ref{eq:dGtu0}) 
it is found that

\begin{equation}
G_t(u;\eta=1) = F(u) (N_0+t)^{-[2m(1-u)+1]} + \frac{1}{2m(1-u) + 1},
\label{eq:Gtu1C}
\end{equation}
 
\noindent
where $F(u)$ is a yet unknown function of $u$ that does not depend on time.
Inserting $t=0$ into Eq. (\ref{eq:Gtu1C}), we obtain

\begin{equation}
G_0(u) = F(u) (N_0)^{-[2m(1-u)+1]} + \frac{1}{2m(1-u) + 1}.
\label{eq:Gtu2C}
\end{equation}

\noindent
Extracting $F(u)$ from Eq. (\ref{eq:Gtu2C}) and inserting it back into Eq. (\ref{eq:Gtu1C}),
we obtain

\begin{equation}
G_t(u;\eta=1) = \beta_t^{ 2m(1-u)+1 } G_0(u)
+ \left[ 1 - \beta_t^{  2m(1-u)+1  } \right]
\frac{1}{ 2m(1-u) + 1 },
\label{eq:Gtu}
\end{equation}
 
\noindent
where

\begin{equation}
\beta_t = \left( 1 + \frac{t}{N_0} \right)^{-1}.
\label{eq:a_t}
\end{equation}

In the long time limit, the generating function converges towards
a steady state of the form

\begin{equation}
G_{\rm st}(u;\eta=1) = 
\frac{1}{ 2m(1-u) + 1 }.
\label{eq:Gtulong}
\end{equation}

\noindent
Expanding Eq. (\ref{eq:Gtulong}) in powers of $u$, we obtain
the steady state degree distribution 

\begin{equation}
P_{\rm st}(k;\eta=1) = \frac{1}{2m+1} \left( \frac{2m}{2m+1} \right)^k,
\label{eq:Pketa1}
\end{equation}

\noindent
which is an exponential distribution.
The mean of the distribution $P_{\rm st}(k;\eta=1)$ is given by

\begin{equation}
\langle K \rangle_{\rm st}(\eta=1) = 2m,
\end{equation}

\noindent
and its variance is given by

\begin{equation}
{\rm Var}_{\rm st}(K;\eta=1) =  2m(2m+1).
\end{equation}

The time dependent degree distribution is obtained by
expanding the right hand side of Eq. (\ref{eq:Gtu}) in powers of $u$.
It yields

\begin{eqnarray}
P_t(k;\eta=1) &=&
\beta_t^{ 2m+1 }  P_0(k)
\nonumber \\
&+& 
\beta_t^{ 2m+1 } 
\sum_{i=1}^{k}  
\frac{ \left[ 2m \ln \beta_t  \right]^i }{i!}
\left[ P_0(k-i) - P_{\rm st}(k-i; \eta=1) \right]
\nonumber \\
&+& \left[ 1 - \beta_t^{ 2m+1 } \right] P_{\rm st}(k; \eta=1).
\end{eqnarray}

\noindent
The mean degree can be obtained from Eq. (\ref{eq:<K>t}),
where $G_t(u;\eta=1)$ is taken from Eq. (\ref{eq:Gtu}).
It is given by

\begin{equation}
\langle K \rangle_t (\eta=1) = \beta_t \langle K \rangle_0 
+  (1-\beta_t) 2m.
\end{equation}
 
To obtain the variance ${\rm Var}_t(K)$
we use the cumulant generating function, which is given by

\begin{equation}
F_t(x;\eta=1) = \ln G_t(e^x; \eta=1).
\label{eq:Ftxeta}
\end{equation}

\noindent
The variance is obtained from

\begin{equation}
{\rm Var}_t(K;\eta=1) = 
\frac{d^2}{dx^2} F_t(x;\eta=1) \bigg\vert_{x=0}.
\label{eq:varketa1}
\end{equation}

\noindent
Inserting $F_t(x;\eta=1)$ from Eq. (\ref{eq:Ftxeta}) into Eq. (\ref{eq:varketa1}),
one finds that

\begin{eqnarray}
{\rm Var}_t(K;\eta=1) &=&
\beta_t {\rm Var}_0(K;\eta=1)
\nonumber \\
&+& (1-\beta_t) {\rm Var}_{\rm st}(K;\eta=1)
\nonumber \\
&+& \beta_t (1-\beta_t)
\left[ \langle K \rangle_0 (\eta=1) - \langle K \rangle_{\rm st}(\eta=1) \right]^2
\nonumber \\
&-& 4 m \beta_t \ln \beta_t
\left[ \langle K \rangle_0(\eta=1) - \langle K \rangle_{\rm st}(\eta=1) \right].
\end{eqnarray}

%\clearpage
%\newpage

\end{document}